\documentstyle[12pt]{article}
\newcommand{\be}{\begin{equation}} \newcommand{\ee}{\end{equation}}

\newcommand{\Tr}{{\rm{Tr}}} \date{} 
\renewcommand{\theequation}{\arabic{section}.\arabic{equation}}
 
\begin{document} \begin{titlepage} \begin{flushright} HD--THEP--94--1
\end{flushright} \quad\\ \vspace{1.8cm} \begin{center} {\bf\LARGE
EVOLUTION EQUATIONS}\\ \bigskip {\bf \LARGE FOR THE QUARK-MESON
TRANSITION}\\ \vspace{1cm} U. Ellwanger\footnote{\footnotesize
Supported by a DFG Heisenberg fellowship, e-mail: I96 at
VM.URZ.UNI-HEIDELBERG.DE}\\ \bigskip C.
Wetterich\footnote{\footnotesize e-mail: T09 at
VM.URZ.UNI-HEIDELBERG.DE}\\ \bigskip Institut  f\"ur Theoretische
Physik\\ Universit\"at Heidelberg\\ Philosophenweg 16, D-69120
Heidelberg\\ \vspace{3cm} {\bf Abstract:}\\ \parbox[t]{\textwidth}{
Evolution equations describe the effect of consecutively integrating
out all quantum fluctuations with momenta larger than some infrared
cutoff scale $k$. We develop a formalism for the introduction of
collective degrees of freedom at some  intermediate scale and derive
the corresponding evolution equations. This allows to account for the
appearance of bound states at some characteristic length scale as, for
example, the mesons in QCD.  The vacuum properties, including
condensates of composite operators, can be directly infered from the
effective action for $k\to0$. We compute in a simple QCD-motivated
model with four-quark interactions the chiral condensate $\bar\psi\psi$
and the effective action for pions including $f_\pi$. A full treatment
of QCD along these lines seems feasible but still requires substantial
work.} \end{center}\end{titlepage} \newpage

\section{Introduction} \setcounter{equation}{0} A system with many
degrees of freedom or a field theory is often described  by different
relevant excitations at different length scales. A famous example in
particle physics is the theory of strong interactions, QCD. At short
distances (on scales below 1 fm) the relevant degrees of  freedom are
quarks and gluons whose interactions are well described by perturbative
QCD \cite{1}. The particles which are observed at large scales,
however, are mesons and hadrons. The interaction of the pseudoscalar
mesons is well modeled by chiral perturbation theory \cite{2}. The
corresponding nonlinear $\sigma$-model shares the flavour symmetries of
perturbative QCD and it is believed that its free phenomenological
parameters can ultimately be computed from the QCD Lagrangian. In
practice, the transition from one set of degrees of  freedom (quarks
and gluons) to another (mesons and hadrons) is a difficult task. Its
solution would greatly enhance the predictivity of QCD, since the large
number of parameters characterizing the meson masses and their low
energy interactions would be reduced to the one free mass scale present
in QCD plus the values of the short distance quark masses. In this
paper we will propose an approach for a description of the
``quark-meson transition'' which is formulated in continuous
spacetime. It is sufficiently general to apply, in principle, to all
problems where the relevant degrees of freedom depend on the length
scale.

Our approach is based on the concept of the effective average action
$\Gamma_k [\varphi]$. $\Gamma_k[\varphi]$ is obtained by integrating
out all modes of  the quantum field with momenta larger than an
infrared cutoff scale, $q^2>k^2$. For $k\to0$ the infrared cutoff is
removed and $\Gamma_k$ becomes  the generating functional of the 1PI
Green functions. In particular the properties of the vacuum and of
excitations around the vacuum can directly be read off from
$\Gamma_{k\to0}$. For nonvanishing $k$ the integration of quantum
fluctuations is only partial. It does not yet include the effects of
modes with $q^2<k^2$. For $k\to\infty$ (or $k$ equal to some
ultraviolet cutoff $\Lambda$) the effective average action equals the
classical action. Knowledge of the $k$-dependence of $\Gamma_k$
therefore allows to interpolate from the classical action to the
effective action, including consecutively more and more quantum
fluctuations as the scale $k$ is lowered. The scale dependence of the
effective average action is described by an exact evolution
equation\footnote{In the present form this equation was first obtained
\cite{3}  as a renormalization-group improved one-loop equation for the
average action. It was later proven by simple manipulation of a
functional integral with an infrared cutoff that eq. (1.1) is an exact
nonperturbative equation \cite{4}. The equivalence of (\ref{1.1}) with
Polchinski's version of an exact renormalization group equation
\cite{5} was established by means of a Legendre transform \cite{6,7}.
(Note that the ``ultraviolet cutoff action'' described by \cite{5} does
not become the effective action as the cutoff goes to zero, but   the
generating functional of one-particle reducible Green functions with
sources replaced by specific functionals of the fields \cite{8,9,10}.
A detailed discussion and comparison can be found in \cite{11}.) JThe
intentions of Polchinski as well as subsequent authors \cite{6,8,12,13}
were mainly aimed towards proofs of perturbative renormalizability.
Many earlier versions of exact renormalization group or  evolution
equations exist in statistical mechanics \cite{14}. Equation
(\ref{1.1}) is surely equivalent to them. Its solutions for $k\to0$ are
expected to be solutions to the Schwinger-Dyson equation \cite{15}.}
($t=\ln k$) \be\label{1.1} \frac{\partial}{\partial
t}\Gamma_k[\varphi]=\frac{1}{2}{\rm Tr}\left\lbrace
\left(\Gamma_k^{(2)}[\varphi]+R_k\right)^{-1}\frac{\partial
R_k}{\partial t} \right\rbrace.\ee Here the trace runs typically over
momenta and internal indices, and  $\Gamma^{(2)}$ denotes the second
functional derivative with respect to the fields $\varphi$. Within a
loopwise expansion, $\Gamma_k=\Gamma_{(0)k}+ \Gamma_{(1)k}+...$, eq.
(\ref{1.1}) has the iterative solution  \be\label{1.2}
\Gamma_k=\Gamma_{(0)k}+\frac{1}{2} \ln {\rm
Det}(\Gamma_{(0)k}^{(2)}+R_k)+...\ee where the dots denote
$k$-independent counter terms and higher-order terms. Thus equation
(\ref{1.1}) easily reproduces the familiar one-loop result, but the
equation is actually exact to all loop orders. The additional infrared
cutoff term $R_k$ in the inverse propagator suppresses the propagation
of modes with $q^2<k^2$.  The appearance of $\Gamma_k^{(2)}$ on the
right-hand side turns equation (\ref{1.1}) into a complicated partial
differential equation for infinitely many variables (typically $k$ and
$\varphi(q)$, the Fourier modes of the field $\varphi$). Alternatively,
equation (\ref{1.1}) can be viewed as an infinite system of coupled
nonlinear differential equations for the flow of the infinitely many
couplings needed to parametrize the most general form of $\Gamma_k$.
In most cases it is impossible to find exact solutions of this equation
and its  practical use may therefore be questioned. The particularly
simple ``one-loop'' form of (\ref{1.1}), the simple interpretation of
$\Gamma_k$ both for $k\to0$ and $k\to\infty$ and the representation of
$\Gamma_k$ in terms of a functional integral with constraint \cite{16}
allow, however, an educated guess on the relevant variables for a given
theory. This permits the ``truncation''  of $\Gamma_k$ with an ansatz
containing only a finite number of couplings. The existence of useful
nonperturbative truncations seems to us the main practical advantage of
(\ref{1.1}) compared to earlier versions of exact renormalization group
equations \cite{14,5},
 whose approximative solutions in general involved again usual
perturbative, large $N$ or $\epsilon$-expansions. In fact, the formal
exactness of the evolution equation constitutes by itself not yet any
calculational progress compared to more standard perturbative expansion
techniques. Only the existence of nonperturbative approximation schemes
for its solutions can provide a key for the exploration of
nonperturbative physics.

Despite its close ressemblance to a one-loop equation the evolution
equation (\ref{1.1}) still contains the full nonperturbative content of
a theory. We observe that  the r.h.s. of (\ref{1.1}) is both infrared
and ultraviolet finite if an appropriate cutoff $R_k$ is chosen, as for
example \be\label{1.3}
R_k=\frac{q^2\exp(-q^2/k^2)}{1-\exp(-q^2/k^2)}\ee \be\label{1.4}
\lim_{q^2\to0}R_k=k^2\ee No need for an additional specification of an
ultraviolet  regularization arises in this case. This is replaced by
the necessary specification of $\Gamma_k$ for some sufficiently large
scale $k=\Lambda$. (Here $\Lambda$ may be associated with a physical
ultraviolet cutoff in the sense that all quantum fluctuations with
$q^2>\Lambda^2$ are already included in $\Gamma_\Lambda$.) A solution
of the exact evolution equation for $k\to0$ with initial condition
specified by the form of $\Gamma_\Lambda$ yields all Green functions
of the theory. The possibility to extract useful nonperturbative
information even  from approximative  solutions of eq. (\ref{1.1}) has
already been demonstrated  by the successful use of nonperturbative
``truncations'': For a $SO(N)$ symmetric scalar field theory the phase
structure in arbitrary dimension $d$ has been  correctly reproduced
\cite{3}, including the Kosterlitz-Thouless phase transition \cite{17}
 for $d=2,N=2$. Critical exponents of the three-dimensional theory have
been computed with a few percent  accuracy \cite{18}.  The high
temperature phase transition in the four-dimensional  theory was
established to be of the second order \cite{19} in contrast to
indications of  earlier perturbative results. (For more related
nonperturbative work on scalar  field theories see refs.~\cite{20}.) Of
direct relevance to the present  work is the description of bound
states \cite{7} through properties of $\Gamma_k$: A nonperturbative
study of the momentum dependence of the four-point function shows the
appearance of a pole in the $s$-channel from which mass and wave
function of the bound state can be extracted. These first encouraging
results for the description of nonperturbative physics raise the
challenge: Can one tackle with these methods the most important
nonperturbative problem in field theory --- the behaviour of QCD for
low momenta?

One problem for such a project is immediately apparent: For large $k$
$(k\gg 1\quad GeV$) the theory is well described in terms of quarks and
gluons and $\Gamma_k$ should take a relatively simple form in terms of
these fields --- typically dominated by gauge-invariant kinetic terms
and mass terms for the quarks. Due to the growing gauge coupling one
expects that the form of $\Gamma_k$ gets more and more complicated as
$k$ comes close to typical scales where strong  interaction phenomena
manifest themselves, say at $k$ around 1 GeV. At these scales effective
four-quark couplings or terms $\sim (F_{\mu\nu}F^{\mu\nu})^2$ do no
longer remain small corrections. For $k$ even much smaller than 1 GeV
the dominant excitations are mesons and hadrons. A description of the
meson and  hadron physics in terms of the original quark and gluon
degrees of freedom must be extremely complicated: Bound states result
in a complicated momentum  dependence of the $n$-point functions and
condensates indicate that an expansion of  $\Gamma_k$ around the
configuration of vanishing fields presumably becomes meaningless. One
clearly needs a transition to express $\Gamma_k$ in terms of new
degrees of freedom more adapted to the physics. In this paper we want
to provide the necessary formalism to include in $\Gamma_k$ collective
fields (as appropriate for the mesons and $\bar\psi \psi$ condensates).
We will again derive an exact evolution equation equivalent to
(\ref{1.1}), but now including the collective fields. Even though on
the formal level there is no need to use these new  ``collective field
equations'' the necessity of an appropriate nonperturbative truncation
scheme makes such an approach almost compulsary. The present work
builds on earlier descriptions of collective fields \cite{10} in the
framework of Polchinski's exact renormalization group equations
\cite{5}.

To be more specific, we expand $\Gamma_k[\varphi]$ in powers of the
``fundamental'' field $\varphi$, with $k$-dependent one-particle
irreducible $n$ point functions $\Gamma_k^{(n)}$ as coefficients. (In
the QCD example $\varphi$ stands for quark and gluon fields.) The
$n$-point functions $\Gamma _k^{(n)}$ will  possibly have a complicated
dependence on the external momenta. Exact flow equations for each of
them can be derived by taking appropriate functional derivatives of
(\ref{1.1}). Let us, for simplicity, consider two-particle bound states
or condensates  in the following: Informations on possible two-particle
bound states are  contained in the four-point function $\Gamma^{(4)}$
and are usually extracted by means of the Bethe-Salpeter equation
\cite{21}. In our approach we follow the $k$-dependent four-point
function $\Gamma_k^{(4)}$ from large towards smaller values of $k$. If
the theory under consideration possesses a two-particle bound state in
the channel $\varphi_1\varphi_2\to\varphi_3\varphi_4$,
$\Gamma_k^{(4)}$ will develop a pole in the momentum variable
$s=(p_1+p_2)^2$ for small enough $k$.\footnote{Even though we work in
Euclidean space where the evolution equation involves $\Gamma^{(n)}$
with all $p_i^2\geq 0$ and $s\geq0$, the final expressions for the
Green functions can be analytically continued to Minkowski space and
$s<0$.} In this case
 $\Gamma_k^{(4)}$ assumes the form \be\label{1.5}
\Gamma_k^{(4)}(p_1,...p_4)\sim-g(p_1,p_2)\tilde G(s)g
(p_3,p_4)+\Gamma_k^{\prime(4)}(p_1,...,p_4)\ee where $g(p_1,p_2)$
corresponds to the (amputated) Bethe-Salpeter wave function and $\tilde
G(s)$ to the bound state propagator approaching a pole-like dependence
on $s$ for small enough $k$. At the pole  $\Gamma_k^{\prime(4)}(p_i)$
remains a bounded function of the momenta $p_1,...,p_4$. In ref.
\cite{7}, where the suitably truncated flow equation for
$\Gamma_k^{(4)}$ was integrated numerically for different models, the
emergence of the structure described in (\ref{1.5}) was indeed
observed. It was demonstrated that both the wave function $g$ and the
propagator $\tilde G(s)$ near the pole can be reliably extracted from
$\Gamma_k^{(4)}$.

We associate the bound state with a composite operator of the type
\be\label{1.6} O(q)=\int \frac{d^4p_1}{(2\pi)^4}\frac{d^4p_2}
{(2\pi)^4}g(p_1,p_2)
\varphi^\dagger(-p_1)\varphi(p_2)(2\pi)^4\delta^4(q-p_1-p_2)\ee A
collective field $\sigma$ is then introduced for the composite operator
$O$. (In our QCD model $\varphi$ stands for the quarks and $\sigma$
denotes the  mesons of the linear $\sigma$-model.) Since the
collective field $\sigma$ associated with the bound state may also
develop a nonvanishing vacuum expectation value, it would  be desirable
to study the complete effective potential of the field $\sigma$ and not
just its part quadratic  in $\sigma$.  (The latter could be obtained
from $\tilde G(s)$ at $s=0$.) This is one of the main motivations for
the formalism to be developed in the next sections. This formalism
deals with the effective average action $\Gamma_k[\varphi,\sigma]$
involving one (or more) collective fields $\sigma$. It requires, at a
certain scale $k_\varphi$, to replace $\Gamma_{k_\varphi}[\varphi]$ by
(cf. eq. (\ref{5.11})) \be\label{1.7}
\Gamma_{k_\varphi}[\varphi,\sigma]=\Gamma_{k_\varphi}[\varphi]
+\frac{1}{2}O^\dagger[\varphi]\tilde G O[\varphi]-\sigma^\dagger
O[\varphi]+ \frac{1}{2}\sigma^\dagger\tilde G^{-1}\sigma\ee The
complete effective action for both fundamental and collective fields
$\Gamma[\varphi,\sigma]$ is then obtained after evolving
$\Gamma_k[\varphi,\sigma]$ from $k=k_\varphi$ down to $k=0$ with the
help of a modified flow equation discussed in sects. 3, 4.

If the four-point function $\Gamma_k^{(4)}$ has, at a scale  $k\sim
k_\varphi$, developed a pole-like structure as in (\ref{1.5}),  a
corresponding term will appear  in the part quartic in $\varphi$ in
$\Gamma_{k_\varphi}[\varphi]$ on the r.h.s. of (\ref{1.7}). The
operator $O[\varphi]$ can then be chosen appropriately such that the
pole-like structure in  $\Gamma_{k_\varphi}^{(4)}$ is cancelled by the
second term on the r.h.s. of  (\ref{1.7}). With this choice the
operator $O[\varphi]$ contains the two-particle wave function of
$g(p_1,p_2)$ of eq. (\ref{1.5}). After the cancellation of the
pole-like structure in $\Gamma^{(4)}_k$ the evolution of
$\Gamma_k[\varphi,\sigma]$ down to $k=0$ can proceed with new "initial
condition" $\Gamma_{k_\varphi}^{(4)}
\simeq\Gamma_{k_\varphi}^{\prime(4)}$. The remaining part of the four
point function stays then bounded (unless further bound states appear).
In this way it  becomes apparent how the introduction  of the
collective field $\sigma$ can remove a dominant strong interaction with
a complicated momentum dependence from the Green functions of the
fundamental fields. The price to pay is an effective average action
with more fields and therefore more possible couplings. Even though at
the scale $k_\varphi$ the effective average action starts only with
terms linear and quadratic in the collective field $\sigma$ as in
eq.(\ref{1.7}), this property will not be  preserved for smaller values
of $k$. The flow equation typically generates interaction terms for
$\sigma$ as well. Despite the apparent complication through the
presence of additional fields, the  approximate (truncated) flow
equations may become much simpler. Since we have now explicitly
included the relevant degrees of freedom, the important physics may be
described in terms of fewer couplings than needed for a reliable
description of $\Gamma_k$ in terms of the fundamental fields alone!

This paper is organized as follows:  In sect. 2 we develop the
formalism of the effective average action including collective fields,
and we derive exact flow equations. In sect. 3 we  present the flow
equations in a version where ``fundamental'' and collective fields are
treated on equal footing. Various strategies how to add or remove
effective degrees of freedom in dependence on the characteristic length
scale of the problem are discussed in sect. 4. For the case of QCD this
gives the formalism how to replace quark degrees of freedom by meson
degrees of freedom. In sects. 5 and 6 we finally apply our formalism to
a QCD motivated model with four-quark interactions. We integrate out
the quarks and obtain an effective potential for the linear
$\sigma$-model. This allows a determination of $f_\pi$ and the chiral
condensate $\bar\psi\psi$. Conclusions and prospects for quantitative
improvements of our results are contained in sect. 7.

\section{Exact evolution equations with composite operators  and
collective fields} \setcounter{equation}{0} Our starting point is the
scale-dependent generating functional for the connected Green functions
\footnote{The second index 0 of $W_{k,0}$ refers to an  optional
infrared cutoff for collective fields to be discussed in sect. 3.}
\be\label{2.1} W_{k,0}[J,K]=\ln \int {\cal D}\chi\exp-S_k[\chi]\ee
\be\label{2.2}
S_k[\chi]=S[\chi]+\Delta_kS[\chi]-J^\dagger\chi-K^\dagger\tilde
GO[\chi]-\frac{1}{2} K^\dagger\tilde GK\ee Here  $\chi^\alpha$ are the
(bosonic) quantum fields of the  theory\footnote{The generalization for
fermionic fields is straightforward \cite{22} and will be given in
Appendix C.} with action $S[\chi]$ and $J^*_\alpha$ the  associated
sources with  \be\label{2.3} J^\dagger\chi=J^*_\alpha\chi^\alpha\ee The
indices $\alpha$ include momenta $q^\mu$, possible Lorentz indices for
vector and tensor fields as well as internal indices. (For example,
the action for free scalar fields  in $d$ Euclidean dimensions reads
$S=\frac{1}{2}\chi^\dagger q^2\chi
=\frac{1}{2}\int\frac{d^dq}{(2\pi)^d}\chi^*_a(q)(q_\mu
q^\mu)\chi^a(q)$. In this  notation functional derivatives obey
$\delta\chi^a(q)/\delta\chi^b(q')=\delta^a_b \delta(q,q')$ where
$\delta(q,q')=(2\pi)^d\delta^d(q-q')$ such that ${\rm Tr}_q\delta
(q,q')=1.$)

Let us neglect first the last two terms in $S_k$ which involve $K$.
Then the only modification of the standard definition of the generating
function $W[J]$ is the addition of an infrared cutoff which is
quadratic in $\chi$: \be\label{2.4}
\Delta_kS[\chi]=\frac{1}{2}\chi^\dagger R_k\chi\ee The matrix
$(R_k)^\alpha_\beta$ will be chosen such that it suppresses the
contribution of fluctuations with  momenta $q^2\ll k^2$ in the
functional integral (\ref{2.1}). We require that the infrared cutoff
$\Delta_kS$ preserves all symmetries of the theory. A typical
choice\footnote{For  precision calculations it is convenient to
incorporate in $R_k(q)$ appropriate wave function renormalization
constants and possibly also mass terms.} for $R_k$ for the case of
scalar fields with internal indices $a$, $b$ is \be\label{2.5}
(R_k)^a_{\ b}(q,q')=R_k(q)\delta^a_b\delta(q,q')\ee \be\label{2.5a}
R_k(q^2)=\frac{q^2\exp(-q^2/k^2)}{1-\exp(-q^2/k^2)} \ee For $q^2\ll
k^2$ one finds $R_k\approx k^2$ and the infrared cutoff scale $k$ acts
like an additional mass. On the other side $R_k$ is exponentially small
for $q^2\gg k^2$ and does therefore not affect the functional
integration of  modes with $q^2\gg k^2$. If desired, however,
$R_k(q^2)$ could also be chosen such  that it incorporates an UV cutoff
$\Lambda$. Then  only fluctuations with $q^2<\Lambda^2$ are included in
the functional integral (see sect. 5). A very intuitive picture arises
if we combine the classical kinetic term from $S[\chi]$ with
$\Delta_kS$ which leads to a lowest order  propagator $G_k^{(0)}(q)$ of
the  form \footnote{If $R_k(q^2)$ involves an UV cutoff $\Lambda$, it
can be chosen such that $G^{(0)}_{k,\Lambda}$ has the form
$G_{k,\Lambda}^{(0)}= ((\exp-q^2/\Lambda^2)-\exp(-q^2/k^2))/q^2$.}
\be\label{2.6a}
G_k^{(0)}(q)=(q^2+R_k)^{-1}=\frac{1-\exp(-q^2/k^2)}{q^2}\ee We see that
the propagation of modes with $q^2\gg k^2$ remains essentially
unchanged whereas modes with $q^2\ll k^2$ cease to propagate.

We observe that for $k\to 0$ and $K=0$ the functional $W_{k,0}$
approaches the standard definition of $W[J]$
\begin{eqnarray}\label{2.6} &&\lim_{k\to0}\Delta_kS=0\nonumber\\
&&\lim_{k\to0}W_{k,0}[J,0]=W[J]\end{eqnarray} Performing a Legendre
transformation of $W_{k,0}$ for $K=0$ and subtracting the  infrared
cutoff term yields the effective average action $\Gamma_{k,0}[\varphi]$
which becomes the generating functional for the 1PI Green functions for
$k\to0$ \cite{4} \begin{eqnarray}\label{2.7}
\Gamma_{k,0}[\varphi]&=&-W_{k,0}[J,0]+J^\dagger\varphi-\frac{1}{2}
\varphi^\dagger R_k\varphi\nonumber\\
&=&\tilde\Gamma_{k,0}[\varphi]-\frac{1}{2}\varphi^\dagger
R_k\varphi\nonumber\\ \varphi^\alpha&=&\frac{\delta
W_{k,0}[J,0]}{\delta J^*_\alpha}\end{eqnarray} The dependence of
$\Gamma_{k,0}$ on the scale $k$ is described by the exact evolution
equation (\ref{1.1}) mentioned in the introduction.

We now want to consider composite operators $O^i[\chi]$ for which we
have introduced  appropriate sources $K^*_i$ in (\ref{2.1}), with
\be\label{2.8} K^\dagger\tilde GO[\chi]=K^*_i\tilde G^i_{\
j}O^j[\chi]\ee Here $\tilde G$ typically plays the role of a propagator
with an appropriate dimensional factor such that the sources have a
standard normalization. In momentum space $\tilde G$ is a function of
$q^2$. For simplicity of notation it is often convenient to use
rescaled operators \be\label{2.8a} \tilde O[\chi]=\tilde G O[\chi]\ee
such that the coupling to the sources takes the standard form
$K^\dagger\tilde O$.  We also have used the freedom \cite{23,21} to add
in (\ref{2.2}) a field-independent term quadratic in $K$ with a matrix
$\tilde G^i_{\ j}$.  The role of this term will become more apparent
later. The generating functional for the connected Green functions for
the fields $\chi$ as well as for the composite operator obtains again
in the limit $k\to0$ \be\label{2.9} \lim_{k\to0}W_{k,0}[J,K]=W[J,K]\ee
For example, the expectation value of the operator $\tilde O^i$ reads
\be\label{2.10} <\tilde O^i>=\frac{\delta W}{\delta
K^*_i}_{|J=0,K=0}\ee and the connected two-point function
is\footnote{We have chosen a normalization appropriate for real
composite operators where $\tilde O^*_i$ and $\tilde O_i$  are not
independent. Our conventions also cover complex fields and operators
if index summations over internal indices include negative values
\cite{4}.} \be\label{2.11} <\tilde O^i\tilde O_j^*>-<\tilde
O^i><\tilde O_j^*>=\frac{\delta^2W}{\delta K^*_i \delta
K^j}_{|J=0,K=0}-\tilde G^i_{\ j}\ee Knowledge of the functional
$W[J,K]$ therefore allows the computation of various  ``condensates''
and correlations between ``condensates''.

More generally, we will define for an arbitrary scale $k$ and arbitrary
source $J,K$ classical fields $\varphi^\alpha$ and $\sigma^i$,
\begin{eqnarray}\label{2.12} \frac{\delta W_{k,0}}{\delta
J^*_\alpha}&=&\varphi^\alpha\nonumber\\ \frac{\delta W_{k,0}}{\delta
K^*_i}&=&\sigma^i,\end{eqnarray} where the composite or collective
field $\sigma^i$ represents the composite  operator $\tilde O_i$. We
combine the fundamental and composite fields and their associated
sources in single vectors \begin{eqnarray}\label{2.13}
j^*_m&=&(J^*_\alpha,K^*_i)\nonumber\\
\psi^m&=&(\varphi^\alpha,\sigma^i)\nonumber\\ \frac{\delta
W_{k,0}}{\delta j^*_m}&=&\psi^m\end{eqnarray} The generalized two-point
function reads then \be\label{2.14} G^m_{\ n}=\frac{\delta^2
W_{k,0}}{\delta j^*_m\delta j^n}\ee We now perform a Legendre transform
with respect to both ``fundamental'' and ``composite'' sources $J$ and
$K$ \be\label{2.15}
\tilde\Gamma_{k,0}[\psi]+W_{k,0}[j]-j^\dagger\psi=0\ee \be\label{2.16}
\frac{\delta\tilde\Gamma_{k,0}}{\delta\psi^m}=j^*_m\ee \be\label{2.17}
G^m_{\ n}\frac{\delta^2\tilde\Gamma_k}{\delta\psi^*_n\delta\psi^p}
=\delta_p^m\ee In particular, the values of $\varphi$ and $\sigma$
extremizing $\tilde\Gamma_{0,0}$ determine the expectation values of
$\chi$ and $\tilde O[\chi]$ and the second functional derivative of
$\tilde\Gamma_{0,0}$ at the minimum gives the exact inverse
propagator.

The exact evolution equation or flow equation describing the dependence
of $\tilde\Gamma_{k,0}$ on the infrared cutoff scale $k$ is now easily
obtained. We notice that the only dependence of $W_{k,0}$ on $k$ arises
through $\Delta_kS$ which is quadratic in $\chi$. The derivative of
$W_{k,0}$ with respect to $t=\ln k$ can therefore be expressed in terms
of the Green function $G$ \begin{eqnarray}\label{2.18}
\frac{\partial}{\partial
t}\tilde\Gamma_{k,0_{|\psi}}&=&-\frac{\partial}{\partial t}
W_{k,0_{|j}} =\frac{\partial}{\partial t}<\Delta_kS>\nonumber\\
&=&\frac{1}{2}\frac{\partial}{\partial t}(R_k)^\alpha_{\
\beta}<\chi^*_\alpha \chi^\beta>\nonumber\\
&=&\frac{1}{2}\left(\frac{\partial R_k}{\partial t}\right)^\alpha_{\
\beta}(G^\beta_{\  \alpha}+<\chi^\beta><\chi_\alpha^*>)\end{eqnarray}
Subtracting from $\tilde\Gamma_{k,0}$ the infrared cutoff
\be\label{2.19} \Gamma_k=\tilde\Gamma_k-\frac{1}{2}\varphi^\dagger
R_k\varphi\ee and using the identity (\ref{2.17}) gives the final form
of the exact evolution equation for the effective average action
$\Gamma_k$ including  composite fields \be\label{2.20}
\frac{\partial}{\partial t}\Gamma_{k,0}=\frac{1}{2}{\rm Tr}
\left\lbrace\frac{\partial R_k}{\partial
t}(\Gamma^{(2)}_{k,0}+R_k)^{-1}\right\rbrace \ee It expresses the scale
dependence of $\Gamma_{k,0}$ in terms of its second functional
derivative \be\label{2.21} \left(\Gamma^{(2)}_{k,0}\right)^m_{\ \
n}=\frac{\delta^2\Gamma_{k,0}}{\delta \psi ^*_m\delta\psi^n}\ee We have
extended in (\ref{2.20}) the matrix $R_k$ to act on vectors $\psi$,
with  $(R_k)^i_{\ j}=0,(R_k)^i_{\ \alpha}=0, (R_k)^\alpha_{\ i}=0$.
Only the $(\alpha,\beta)$ components of $R_k$ (the ones corresponding
to the fundamental fields $\varphi^\alpha$) depend on $k$ in this
formulation. Even though only the $\alpha,\beta$ components of
$(\Gamma^{(2)}_{k,0}+R_k)^{-1}$ contribute therefore in (\ref{2.20}),
the $(i,j)$ and $(i,\alpha)$ components of $\Gamma^{(2)}_{k,0}$ are
relevant since they enter in forming the inverse of
$\Gamma_{k,0}^{(2)}+R_k$.

For $k\to\infty$ $\Gamma_{k,0}[\varphi,\sigma]$ approaches
\be\label{2.25b} \Gamma_{k\to\infty,0}[\varphi,\sigma]=S[\varphi]+
\frac{1}{2} O^\dagger[\varphi]\tilde G O[\varphi] -\sigma^\dagger
O[\varphi] +\frac{1}{2}\sigma^\dagger\tilde G^{-1}\sigma.\ee   In a
theory with an ultraviolet cutoff $\Lambda$  (cf. footnote 7), the
average action becomes equal to the r.h.s. of eq. (\ref{2.25b}) for
$k=\Lambda$. Solving the evolution equation (\ref{2.20}) with this
initial condition interpolates from the short distance physics
described by $\Gamma_\Lambda$ to the physics at longer distances
described by $\Gamma_k,k<\Lambda$. In this process the quantum
fluctuations with momenta $\Lambda^2>q^2>k^2$ are integrated out.
$\Gamma_k$ for $k\to0$ therefore amounts to the complete effective
action of the  quantum field theory.

In summary, we have presented here an extended form of the exact
evolution equation (\ref{1.1}). It uses additional composite fields
$\sigma$. JA solution of $\Gamma_k[\varphi,\sigma]$ for $k\to0$ encodes
all  information of the original procedure, namely all 1PI Green
functions for the fundamental fields $\chi$. In addition, it contains
the  complete information on expectation values of composite operators
$\tilde O[\chi]$ and their correlations. At first sight the use  of an
enlarged matrix $\Gamma_{k,0}^{(2)}$ may seem to be an additional
complication which is the price for gaining additional information
about Green functions for  composite operators.  It may  happen,
however, that physically important pieces in $\Gamma_k$ have a rather
complicated form when expressed in terms of the ``fundamental fields''
$\varphi$, but become simple once they are written in terms of
composite operators. In this case the use of more fields
$(\varphi,\sigma)$ may result in a considerable simplification of
calculations. This applies in particular to models with condensates of
collective fields and/or bound states. For the theory of strong
interactions we may associate $\varphi$ with the gluon and quark
degrees of freedom and $\sigma$ with operators like
$\bar\psi\psi,F_{\mu\nu}F^{\mu\nu}$ or mesons, hadrons, and glueballs.
The low energy part of the effective action is expected to become much
simpler when expressed in terms of $\sigma$ rather than of $\varphi$.

\section{The two-field formalism} \setcounter{equation}{0} In a variety
of physical situations there is no clear distinction between
fundamental  and composite fields. This applies, for example, if one
deals with processes with typical momentum transfers much smaller than
the mass scale charcteristic for the structure of the bound state. The
structure cannot be resolved in this case, and the bound state appears
as an ordinary particle. An example are versions of the  standard model
where the Higgs scalar is a bound state. In situations of this type one
would like to treat fundamental fields and composite fields on a
completely equal footing (at least at low energies) such that standard
field-theoretical methods as perturbation theory can be applied. For
this purpose it is instructive  to write the functional $W_{k,0}[J,K]$
in terms of a functional integration over  both fundamental and
collective degrees of freedom. We use the identity (choosing $\tilde G$
such that $\tilde G^\dagger=\tilde G$) \be\label{3.1} 1=N\int {\cal
D}\rho\exp\left\lbrace-\frac{1}{2}(\rho^\dagger -K^\dagger\tilde
G-O^\dagger[\chi]\tilde G)\tilde G^{-1}(\rho-\tilde GK-\tilde
GO[\chi])\right \rbrace\ee with $N$ an irrelevant constant to be
omitted in the following. Insertion in (\ref{2.1}) yields
\be\label{3.2} W_{k,0}[J,K]=\ln\int D\chi D\rho\exp-S_k[\chi,\rho]\ee
with \begin{eqnarray}\label{2.24}
&&S_k[\chi,\rho]=S[\chi]-J^\dagger\chi+\Delta_kS[\chi]\nonumber\\
&&+\frac{1}{2}O^\dagger[\chi]\tilde GO[\chi]-\rho^\dagger O[\chi]
-K^\dagger\rho+\frac{1}{2}\rho^\dagger\tilde G^{-1}\rho.\end{eqnarray}
We see that $W_{k,0}[J,K]$ has now the same form of a theory with
additional fields and interactions. In particular, $S_k$ is linear in
the source $K$ which motivates the term $\sim K^\dagger\tilde GK$ in
the original formulation (\ref{2.2}). Compared to the formulation in
terms of only fundamental fields $\chi$ we have  introduced new fields
$\rho$ with inverse propagator $\tilde G^{-1}$ and interactions  with
the fundamental field $\sim\rho^\dagger O[\chi]$. For the example of an
operator $O$ quadratic in $\chi$ one obtains a cubic interaction as
depicted in fig. 1. One also has an additional term $\sim
O^\dagger\tilde GO$ involving only the fields $\chi$. It has the form
of an exchange of $\rho$ in the tree approximation (fig. 2) and can be
used to cancel terms in $S[\chi]$ with a pole structure as discussed in
the introduction and in more detail in sections 4 and 5.

Inside the one-loop diagram corresponding to the r.h.s. of the flow
equation (\ref{2.20}) also collective fields appear. Their presence is
induced by the inversion of $(\Gamma^{(2)}_{k,0}+R_k)$ in the enlarged
space with indices $(i,\alpha)$, (cf. the discussion following  eq.
(\ref{2.21})).  The  effective propagator $\tilde G$ of the collective
fields $\rho$ does not yet include an infrared regulator; thus it is
possible that the r.h.s. of the flow equation (\ref{2.20}) contains, in
a given truncation, an infrared divergence. This  situation can
especially arise in the case of dynamical symmetry breaking, where some
collective degrees of freedom become massless Goldstone
bosons.\footnote{In practice one  might sometimes consider a truncation
of $\Gamma_{k,0}$ where vertices generating ``internal'' collective
fields are discarded, see sect. 5. Here,  however, we consider the
general case.} In the formulation  (\ref{2.24}) it becomes easy to
introduce an additional infrared cutoff (with scale $\tilde k$) for the
collective field $\rho$. We can generalize $W_{k,0}$ to $W_{k,\tilde
k}$ by adding to $S_k$ (\ref{2.24}) a term \be\label{2.25}
\tilde\Delta_{\tilde k}S[\rho]=\frac{1}{2}\rho^\dagger\tilde R_{\tilde
k}\rho.\ee The matrix $(\tilde R_{\tilde k})^i_{\ j}$ should have
similar  properties as $(R_k)^\alpha_{\ \beta}$. In particular it
should vanish for $\tilde k\to0$, and all eigenvalues should diverge
for $\tilde k\to\infty$ (or $\tilde k\to\Lambda$).  Correspondingly,
$\tilde \Gamma_{k,\tilde k}$ is defined by the analogue of (\ref{2.15})
and $\Gamma_{k,\tilde k}$ reads \be\label{2.26} \Gamma_{k,\tilde
k}=\tilde\Gamma_{k,\tilde k}-\frac{1}{2} \varphi^\dagger
R_k\varphi-\frac{1}{2}\sigma^\dagger\tilde R_{\tilde k}\sigma.\ee We
note \be\label{2.27} \frac{\delta W_{k,\tilde k}}{\delta K_i^{\
*}}=<\rho^i>=\sigma^i\ee with the expectation value evaluated now with
the action  $S_k[\chi,\rho]+\tilde\Delta_{\tilde k}S[\rho]$. The
evolution equation for the $k$-dependence of $\Gamma_{k,\tilde k}$ at
fixed $\tilde k$ has the same form as (\ref{2.20}) with $(R_k)^i_{\ j}$
replaced by $(\tilde R_{\tilde k})^i_{\ j}$,  which is now different
from zero but independent of $k$. (Only the denominator on the r.h.s.
of (\ref{2.20}) changes due to the new definition (\ref{2.26}). As in
the previous section, $(R_k)^\alpha_{\ \beta}$ and $(\tilde R_{\tilde
k})^i_{\ j}$ are combined to an enlarged matrix $R_{k,\tilde k}$.) The
dependence of $\Gamma_{k,\tilde k}$ on  $\tilde t=\ln \tilde k$ can
also be  derived in complete analogy to (\ref{2.20}) \be\label{2.28}
\frac{\partial}{\partial \tilde t}\Gamma_{k,\tilde k}=\frac{1}{2} {\rm
Tr}\left\lbrace\frac{\partial R_{k,\tilde k}}{\partial \tilde t}
(\Gamma_{k,\tilde k}^{(2)}+R_{k,\tilde k})^{-1} \right\rbrace\ee Here
$R_{k,\tilde k}$ acts on vectors $(\varphi,\sigma)$ as defined above
and only $\frac{\partial}{\partial \tilde t}(R_{k,\tilde k})^i_{\
j}=\frac{\partial}{\partial \tilde t}(\tilde R_{\tilde k})^i_{\ j}$  is
different from zero in this  case. In particular, we may identify $k$
and $\tilde k$ and define \be\label{2.29}
\Gamma_k[\varphi,\sigma]=\Gamma_{k,k}[\varphi,\sigma]\ee The evolution
equation for the $k$-dependence of $\Gamma_k$ is again given by
(\ref{2.20}) since \be\label{2.30} \frac{\partial}{\partial
t}\Gamma_k=\frac{\partial}{\partial t}\Gamma_{k,\tilde k |\tilde
k=k}+\frac{\partial}{\partial\tilde t}\Gamma_{k,\tilde k|\tilde k=k}\ee
Now both $(R_k)^\alpha_{\ \beta}$ and $(R_k)^i_{\ j}$  depend on $k$.
The evolution equation for the two sorts of fields $\varphi$ and
$\sigma$ has  exactly the same form as the original equation
(\ref{1.1}).

In the limit $k\to\Lambda,\tilde k\to \Lambda$ ($\Lambda$ may be
infinity)  the functional integrals defining $W_k=W_{k,k}$ and
$\Gamma_k$ are easily solved. The term quadratic in the fields $\sim
R_k$ diverges, the classical approximation becomes exact and one
obtains the same result as in (\ref{2.25b}): \be\label{2.31}
\Gamma_\Lambda[\varphi,\sigma]=S[\varphi]+\frac{1}{2}
O^\dagger[\varphi]\tilde G O[\varphi] -\sigma^\dagger O[\varphi]
+\frac{1}{2}\sigma^\dagger\tilde G^{-1}\sigma\ee The equivalence of the
initial condition (\ref{2.31}) with the original formulation in terms
of  fundamental fields only is readily established by solving the field
equations for $\sigma$ and inserting in $\Gamma_\Lambda$
\begin{eqnarray}\label{2.32}
&&\frac{\delta\Gamma_\Lambda}{\delta\sigma}_{|\sigma_0}=0\nonumber\\
&&\sigma_0=\tilde GO[\varphi]=\tilde O[\varphi]\nonumber\\
&&\Gamma_\Lambda[\varphi,\sigma_0]=S[\varphi]\end{eqnarray}

In the opposite limit $k\to0,\tilde k\to0$ the infrared cutoff term
$\sim R_k$ vanishes and $\Gamma_0$ becomes $\Gamma_{0,0}$ of sect. 2,
the generating functional for the 1PI Green  functions for the fields
$\varphi$ and $\sigma$.  Inserting the solution of the field equation
for $\sigma$ which minimizes $\Gamma_0[\varphi,\sigma]$ \be\label{2.33}
\frac{\delta\Gamma_0}{\delta\sigma}_{|\sigma_0}=0\ee we recover the
generating functional for the 1PI Green functions for the fundamental
field $\chi$ (since $K=0$) \be\label{2.34}
\Gamma[\varphi]=\Gamma_0[\varphi,\sigma_0[\varphi]]\ee The full
generating functional $\Gamma_0[\varphi,\sigma]$ contains all
information on the vacuum expectation of the operator $\tilde O[\chi]$
as well as all $n$-point functions of this operator. The procedure for
extracting these quantities is straightforward and given  explicitly in
appendix A.

\section{Scale-dependent degrees of freedom} \setcounter{equation}{0}
The relevant degrees of freedom often depend on the length scale. For
the example of QCD one would like to describe the short-distance
physics in terms of gluons and  quarks and the long-distance physics in
terms of mesons and hadrons. In  sect. 3 we have introduced the
effective average action $\Gamma_{k,\tilde k}$ depending on two scales
$k$ and $\tilde k$, which will allow us to describe a smooth transition
between different sorts of relevant degrees of freedom. Consider first
the case that below some momentum scale $k_\sigma$ we want to describe
the physics uniquely in terms of the collective fields $\sigma$. For
example, one may want to describe QCD below $k_\sigma\approx 500$ MeV
by a linear or nonlinear $\sigma$-model for the pseudoscalar mesons. In
our approach this can be done very naturally by first fixing the
infrared regulator for the collective fields $\tilde k=k_\sigma$ and
solving the evolution equation for the infrared regulator for the
fundamental fields $k\to0$. The resulting effective action for
composite fields $\Gamma_{\tilde k}[\sigma]$ is given by \be\label{4.1}
\Gamma_{\tilde k}[\sigma]=\Gamma_{0,\tilde k}
[\varphi_0[\sigma],\sigma]\ee where \be\label{4.2}
\frac{\delta\Gamma_{0,\tilde
k}}{\delta\varphi}[\varphi_0[\sigma],\sigma]=0\ee for $\tilde k=\tilde
k_\sigma$. It can be used as a starting point for integrating out the
``mesons'' with momenta $q^2<k^2_\sigma$. The low momentum fluctuations
of the mesons are not yet accounted for in $\Gamma_{\tilde k}[\sigma]$
due to the existence of an effective infrared cutoff (\ref{2.25}). They
will be included by evolving $\Gamma_{\tilde k}[\sigma]$ to
$\Gamma_0[\sigma]$  using the
 evolution equation (\ref{2.28}). We observe that $R$ is now
nonvanishing only for the composite part $(R^\alpha_{\ \beta}=0$ since
$k=0$,  $R^i_{\ j}=(\tilde R_{\tilde k})^i_{\ j}$) and that
$\Gamma_{0,\tilde k}^{(2)}$ becomes block-diagonal in composite and
fundamental fields \be\label{4.3} \frac{\delta^2\Gamma_{0,\tilde
k}}{\delta\sigma\delta\varphi}_{| \varphi_0,\sigma}=0\ee provided
$\varphi_0[\sigma]$ (\ref{5.2}) is independent of $\sigma$. (If the
fundamental fields $\varphi$ assume no vacuum expectation values, the
solutions of eq. (4.2) will simply be $\varphi_0[\sigma]=0.$) Then the
evolution equation (\ref{2.28}) only involves $\Gamma_{\tilde
k}^{(2)}[\sigma]$ on the r.h.s. and has exactly the same form as
(\ref{1.1}), now expressed in  terms of composite fields. This
statement generalizes to an arbitrary dependence of $\varphi_0$ on
$\sigma$ as briefly explained in appendix B. We conclude that the
solution of $\Gamma_{k,\tilde k}$ for $k\to0$ at fixed $\tilde k$ can
be used for a complete replacement of the fundamental variables
$\varphi$ by the composite variables $\sigma$. This procedure amounts
to  integrating out the fluctuations $\varphi$ in a background
$\sigma$.\footnote{The actual calculations of the present paper in
sects. 5,6 will  not involve internal  propagators of the
$\sigma$-field, thus  the introduction of a scale $k_\sigma$ is not
necessary.} The same procedure can actually also be applied to the
problem how to integrate out only one sort of fundamental fields
(without reference to collective degrees of freedom). One may introduce
a variable infrared cutoff with scale $k$ only for the fields to be
integrated out, while the remaining fields have a cutoff with scale
$\Lambda$. Lowering $k$ to zero at fixed $\Lambda$ then only includes
quantum fluctuations of the fields to be integrated out, and
(\ref{4.1}), (\ref{4.2}) with $\tilde k\equiv\Lambda$ give the
effective action for the remaining fields $\sigma$.

Under certain circumstances  it may be convenient to consider both
fundamental and composite degrees of freedom on an equal footing. This
is done most easily by a study of the evolution equation (\ref{2.30})
where $\tilde k=k$. An example are top condensate models at scales much
below the composite scale \cite{24}, which have already been treated by
methods closely related to the present one \cite{25}. Treating the
scalar bound state (the composite Higgs doublet) similar to the
fundamental fermions will reproduce most easily the results of the
perturbative  standard model since for $k=\tilde k$ the fermion and
scalar loops described  by the formal expression (3.9) are treated in
the same way. This is not guaranteed in the (``asymmetric'') version
(2.23) where $\tilde k=0$. Since the standard contribution of the
scalar loops is missing, a relatively complicated truncation of
$\Gamma_{k,0}$ may be needed in order to reproduce even the standard
one-loop result for the running of weak couplings. Another  typical
application of the running of $\Gamma_{k,k}$ is the transition region
between the quark description and meson description in QCD at scales
between 500 MeV and 2 GeV.

Bound states are, in general, characterized by a typical ``composite''
scale. At length scales much shorter than the inverse composite scale,
they play no particular role. For large enough $k$ not only the
introduction of a second  scale $\tilde k$ seems cumbersome, but the
whole formalism with  composite degrees of freedom seems sometimes not
to be very well adapted. One does not want to describe  asymptotic
short-distance QCD by dealing explicitly with mesons! This problem can
be coped with by ``switching on'' the collective degrees of freedom
only for the evolution at $k$ smaller than some scale $k_\varphi$. This
can be achieved  by an appropriate choice of the composite operator
$O[\chi]$. Actually, already the presence of a wave function
$g(p_1,p_2)$ in $O[\varphi]$ (see eq.(1.6)	),
 which decays for large momenta $p^2_i$, will suppress the
contributions of the new terms to the flow of
$\Gamma_k[\varphi,\sigma]$ for large $k^2$.  This effect is much
enhanced by the introduction of exponentially decaying functions of
$p^2_i$ into the operator $O[\varphi]$. Let us represent $O$ as a
functional of the Fourier modes $\chi(q)$ in the form \be\label{4.4}
O[\chi]=\hat O[f_{k_\varphi}(q)\chi(q)]\ee \be\label{4.5}
f_{k_\varphi}(q)=\exp{(-{q^2}/2{k^2_\varphi})}\ee Typically $\hat O$
may be some polynomial of $f_{k_\varphi}(q)\chi(q)$ involving
appropriate wave functions --- see the next section.  More general
forms of $\hat O$ are also allowed, provided it does not  contain an
exponentially growing momentum dependence. The appearance of the
exponential suppression factor $f_{k_\varphi}$ in the coupling between
$\rho$ and $\chi$ (\ref{2.24}) will then  effectively ``switch off''
all effects of the composite fields for  $q^2\gg k^2_\varphi$. Indeed,
for $k>A k_\varphi$ with $A$ sufficiently large  (say $A=10$) all
contributions from composite fields in the evolution equation
(\ref{2.20})  are suppressed. This allows us to compute
$\Gamma_{Ak_\varphi} [\varphi]$ by integrating the evolution equations
for $A k_\varphi<k<\Lambda$ without using the formalism with composite
fields (solving (\ref{1.1})). On the other hand, the modified operator
$O[\varphi]$ still describes appropriately the ``bound state'' for
momenta $p^2_i<k^2_\varphi$ of the fundamental fields. At
$k=Ak_\varphi$ one may switch to the two-field formalism. The initial
terms on the r.h.s. of eqs. (\ref{2.25b}) or  (\ref{2.31}), which
formally have to be introduced for $k\to\infty$ (or $k=\Lambda$), did
not affect the evolution  of $\Gamma_k$ for $k> Ak_\varphi$, and can
thus be introduced at $k= Ak_\varphi$ only: \be\label{4.6}
\Gamma_{Ak_\varphi,\tilde
k}[\varphi,\sigma]=\Gamma_{Ak_\varphi}[\varphi]+\frac{1}{2}
O^\dagger[\varphi]\tilde GO[\varphi] -\sigma^\dagger
O[\varphi]+\frac{1}{2}\sigma^\dagger\tilde G^{-1}\sigma\ee At this
stage one has several options for the infrared cutoff $\tilde k$ for
the collective fields $\sigma$, as described before: One can put it
equal to zero right away, or one can put it equal to $Ak_\varphi$ and
treat it as a variable independent of $k$ subsequently (using eq.
(2.23)), or one can identify it with $k$ both at the starting point
$k=Ak_\varphi$ and concerning the subsequent evolution, using eq.
(3.9). Note that the r.h.s. of eq. (\ref{4.6}) does not depend on the
choice of $\tilde k$; the dependence of $\Gamma_{Ak_\varphi,\tilde k}$
on $\tilde k$
 manifests itselfe only in the form of the evolution equations. The
collective degrees of freedom will now affect the evolution of
$\Gamma_{k,\tilde k}[\varphi,\sigma]$ for $k<Ak_\varphi$ due to the
disappearance of the exponential suppression factor (\ref{4.5}) in the
$\sigma- \varphi$ coupling.

The concrete choice of the operators $O[\varphi]$ and the form of the
propagator $\tilde G$ for the collective fields $\sigma$ depends on the
problem under consideration. For the case of propagating bound states
it has been sketched in the introduction. Let us concentrate in the
following   on a ``two-particle bound state'' showing up in the
four-point function $\Gamma_k^{(4)}$. Our starting point is the
effective action $\Gamma_k[\varphi]$ without collective fields
$\sigma$. After the integration of the corresponding flow equations
(\ref{1.1}) down to a certain  ``bound state'' scale  it is assumed
that a pole-like structure as in (\ref{1.5}) has emerged as part of the
four-point function $\Gamma^{(4)}_k$. It is now desirable to choose the
operator $O[\varphi]$ and the propagator $\tilde G$ such that this
pole-like structure within $\Gamma^{(4)}_k$ gets cancelled; then the
collective field $\sigma$ associated with the operator $O[\varphi]$
corresponds to the bound state degree of freedom generated by the
dynamics of the theory. (Formally no error is made if $O[\varphi]$ is
not chosen appropriately, but then informations on the bound state are
still contained partially in the remaining part  of the four-point
function of the fundamental fields.) An obvious possible choice is to
identify  the ansatz for $\tilde G(q^2)$ in eqs. (\ref{2.2}), (3.1) and
(4.6) with the dynamically generated  $\tilde G(s)$ in eq. (\ref{1.5})
and to take the Fourier components of $O[\varphi]$ as \be\label{4.8}
O(q)=\int
\frac{d^4p_1}{(2\pi)^4}\frac{d^4p_2}{(2\pi)^4}f_{k_\varphi}(p_1)
f_{k_\varphi}(p_2)g(p_1,p_2)\varphi(p_1)\varphi
(p_2)(2\pi)^4\delta^4(p_1+p_2-q)\ee Here the scale $k_\varphi$ should
be somewhat above the scale where a pole-like structure in
$\Gamma^{(4)}_k$ emerges. This ensures that $f_{k_\varphi}$  does not
switch off the collective fields in the momentum range where they are
needed.  On the other hand the presence of $f_{k_\varphi}$ justifies
the introduction of the additional terms in eq. (\ref{4.6}) only at $k=
A k_\varphi$ and not already at $k\to\infty$ (or  $k=\Lambda$) as in
eqs. (2.25), (3.10).  In practice it might also be useful to try
modified forms of the operator $O[\varphi]$ and $\tilde G$ introduced
at $k=Ak_\varphi$; the optimal choice is the one which completely
prevents the appearance of a pole-like structure in $\Gamma_k^{(4)}$
for $k\ll k_\varphi$.

The  methods discussed in this section allow
 for describing smoothly the transition from perturbative QCD to the
nonlinear $\sigma$-model. Variations of the transition scale $A
k_\varphi$ and the different options for the infrared cutoff $\tilde k$
for the collective fields would allow for a variety of checks.
Obviously, much work needs to be done before such a program can be
implemented. We nevertheless will demonstrate the viability of our
method by the investigation of a simplified model in sect. 5. We
finally mention that the restriction of the formalism developed in the
last sections to scalar fields is by no means necessary. The
generalization for the inclusion of (chiral) fermions is presented in
appendix C. Inclusion of gauge fields  requires more thought but seems
not to encounter insurmountable difficulties  \cite{9,13,26}.

\section{A QCD-motivated model} \setcounter{equation}{0} So far we have
developed the general formalism for the description of bound states
and collective fields in the context of the average action in a very
general but  necessarily also rather abstract way. In this section we
want to give an example  which shows how these ideas work in practice.
Our aim is a demonstration of feasibility of our program rather than an
attempt to compute with precision. We nevertheless take a model which
is inspired by QCD. It contains sufficiently many details such that  it
can account for mesons and chiral condensates and can later be extended
to full QCD  (see the discussion in sect. 7).  We work with massless
quarks even though the  introduction of quark mass terms poses  no
particular technical problem.

We start with a theory with only  fermion fields $\psi$.  They
correspond to massless quarks and play the role of the fundamental
fields $\varphi$ of the last sections. We have in mind a version of
QCD, where the gluons have already been integrated out. As a result we
expect that, among others, a (generally nonlocal) four-quark
interaction $\Gamma^{(4)}_\Lambda(p_i)$ has been generated at a scale
$\Lambda\sim$ 1.5 GeV.  Concerning the form of  $\Gamma^{(4)} _\Lambda
(p_i)$  we assume   that for $p^2_i<\Lambda^2$ it depends just on the
Mandelstam variable $t$  such that it corresponds to the sum of a
linear  and a confining potential in the nonrelativistic limit.  We
consider an expansion of the effective action up to  fourth order in
the quark  fields, and we neglect the running of the effective quark
propagator. (Due to chiral symmetry the quarks will remain massless
anyhow; we only neglect the running of the quark wave function
normalization.) Thus we just have to  integrate the evolution equation
for $\Gamma^{(4)}_k$ with the above $\Gamma^{(4)} _\Lambda$ as boundary
condition at $k=\Lambda$. Now we observe indeed that $\Gamma^{(4)}_k$
at some small scale $\sim k_\varphi$ develops a pole-like structure as
in (1.5). It allows us to read off the wave function $g(p_1,p_2)$ and
the collective  field propagator $\tilde G(s)$.

Next we neglect the difference between the bound-state scale,
$k_\varphi$ and $A k_\varphi$, and construct
$\Gamma[\psi,\bar\psi,\sigma]$ at the scale $k_\varphi$ according to
the rule eq. (4.6). (Here we put $f_{k_\varphi}=1$ for simplicity.)
 By this procedure the model is transformed into a linear
$\sigma$-model with momentum dependent  Yukawa couplings between mesons
and quarks. The part of $\Gamma_k$ involving the  scalars starts  at
this scale only with a quadratic term and Yukawa couplings as given by
the r.h.s. of eq. (4.6).  At scales $k$ below $k_ \varphi$ we focus our
attention on the effective potential for the collective field $\sigma$.
It will be generated after integrating out the  quark fields
$\psi,\bar\psi$ with momenta $q$ with $0\leq q^2\leq k^2_\varphi$, due
to the  (momentum dependent) Yukawa coupling. The dominant effect is
the  ``quadratic running'' \cite{16,22} of the scalar mass term. The
fermion fluctuations induce a negative mass term  at the origin
$(\sigma=0)$ and trigger therefore the spontaneous breaking of chiral
symmetry. We neglect contributions to the effective potential due to
internal $\sigma$-lines. This allows to compute the effective potential
 from a simple ``one-loop formula'' (eq. (\ref{6.19}) below) in terms
of the parameters present in
$\Gamma_{k_\varphi}[\psi,\bar\psi,\sigma]$.

To be more precise, we describe the fundamental quark fields by four
component Dirac spinors $\psi^a_i,\bar\psi_a^i$ where the flavour index
$a$ runs from $1... N_f$, $i=1...N_c$ is the colour index and we omit
the spinor indices.  The effective  action contains terms quadratic and
quartic in the quark fields, and is taken to  be invariant under
$SU(N_c)$ and  chiral $U(N_f)_V \otimes U(N_f)_A$ symmetries:
\begin{eqnarray}\label{5.1} \delta_V\psi^a&=&i\theta^a_{V
b}\psi^b,\quad \delta_A\psi^a=i\gamma_5\theta^a_{A b}\psi^b\nonumber\\
\delta_V\bar\psi_a&=&-i\bar\psi_b\theta^b_{V a},\quad
\delta_A\bar\psi_a=i \bar\psi_b\theta_{Aa}^b \gamma_5. \end{eqnarray}
The quadratic part  describes the inverse quark propagator, and a
possible $k$ dependent wave function normalization is put equal to 1
(we contract over the not explicitly written spinor  indices.):
\be\label{5.2} \Gamma_{2,k}=\int\frac{d^4q}{(2\pi)^4}\bar\psi^i_a(q)
q\llap/ \psi^a_i(q) \ee Even in the presence of the vector and axial
vector  symmetries (5.1) a number of different four-quark interactions
can be written down. We restrict ourselves to one particular spin,
flavour and colour structure: \begin{eqnarray}\label{5.3}
&&\Gamma_{4,k}=\frac{1}{2}\int\prod_{l=1}^4\left(\frac{d^4p_l}
{(2\pi)^4}
\right) (2\pi)^4\delta^4(p_1+p_2-p_3-p_4) \lambda_k(p_1,p_2,p_3,p_4)\\
&&\times\{[\bar\psi_a^i(-p_1)\psi^b_i(p_2)][\bar\psi^j_b(p_4)\psi^a_j
(-p_3)]
-[\bar\psi^i_a(-p_1)\gamma_5\psi^b_i(p_2)][\bar\psi^j_b(p_4)\gamma_5
\psi^a_j(-p_3)]\} .\nonumber\end{eqnarray} This structure colour
singlets with nontrivial flavour in the $s$-channel (they will  later
correspond to the mesons) and flavour singlets transforming as colour
singlets and octets  in the $t$-channel (as mediated, for example,  by
a one-gluon exchange). As usual the kinematic variables $s$ and $t$
are defined here by \begin{eqnarray}\label{5.4}
s&=&(p_1+p_2)^2=(p_3+p_4)^2\nonumber\\ t&=&(p_1-p_3)^2=(p_2-p_4)^2.
\end{eqnarray} The contractions of spinor indices are indicated by
brackets, and the invariance under (5.1) is readily checked.

The sum of $\Gamma_{2,k}$ and $\Gamma_{4,k}$ has now to be inserted
into the fermionic version of the flow equation: \be\label{5.5}
\partial_t\Gamma_k=-\Tr\left\{\frac{\partial R_{kF}}{\partial
t}(\Gamma_k^{(2)} +R_{kF})^{-1}\right\}. \ee We use here\footnote{For
this choice $R_k(q)$ does not uniformly approach zero for $k\to 0$. The
divergence for $R_k(q\to 0)$ is of no relevance in the present context}
\be\label{5.6}
R_{kF}(q)=q\llap/\frac{1-exp({-{q^2}/{\Lambda^2}})+exp({-{q^2}/{k^2}})}
{exp({-{q^2}/{\Lambda^2}})-exp({-{q^2}/{k^2}})} \ee such that  the sum
of the terms  $\Gamma_k^{(2)}+R_{kF}$ gives \be\label{5.7}
\frac{q\llap/}{exp(-{q^2}/{\Lambda^2})-exp(-{q^2}/{k^2})}+
O(\bar\psi\psi). \ee We are interested in the terms
$\sim(\bar\psi\psi)^2$ of eq. (5.5), which  contribute to the running
of the coefficient  $\lambda_k(p_1,p_2,p_3,p_4)$ of $\Gamma_{4,k}$.
After a  corresponding expansion of the r.h.s. of eq. (5.5) the
equation describing the  running of $\lambda_k$ assumes the schematic
form shown in fig. 3. Here the lower inner line denotes the
regularized fermionic propagator as derived from the inverse of
$\Gamma_k^{(2)}+R_k$, \be\label{5.8}
G_{2,k}(q)=\frac{exp(-{q^2}/{\Lambda^2})-exp(-{q^2}/{k^2})}{q\llap/},
\ee and the upper crossed line its derivative with respect to $t=\ln
k$. (In fig. 3 we have not shown  a  similar diagram with the cross on
the lower line.) Actually, inserting the ansatz (5.3) for
$\Gamma_{4,k}$ into the r.h.s. of eq. (\ref{5.5}), also different
spin, flavour, and colour structures  than the one of eq. (\ref{5.3})
are generated on the l.h.s. of eq. (\ref{5.5}). They  are put to zero
in our truncation. Since only the retained contribution is
proportional to a combinatorial colour factor  $N_c$ our truncated
evolution equation  becomes exact in the leading order in a $1/N_c$
expansion. After taking the various combinatorial factors into account
and performing the trace over spinor indices, the flow equation  for
$\lambda_k$ becomes \begin{eqnarray}\label{5.9} &\partial_t
\lambda_k(p_1,p_2,p_3,p_4)=\nonumber\\
&-\frac{8N_c}{k^2}\int\frac{d^4q} {(2\pi)^4}
\lambda_k(p_1,p_2,q,-q+p_1+p_2)\lambda_k(q,-q+p_1+p_2,p_3,p_4)
\nonumber\\
&\Bigl[\frac{q^\mu(q-p_1-p_2)_\mu}{q^2}\exp({-{(q-p_1-p_2)^2}/{k^2}})
\left(\exp({-{q^2}/{\Lambda^2}})-\exp({-{q^2}/{k^2}})
\right)\nonumber\\ &+(q\to p_1+p_2-q)  \Bigr]\end{eqnarray}

In the model under consideration this flow equation for $\lambda_k$
will be integrated with a boundary condition for $\lambda_k$ at
$k=\Lambda\simeq$ 1.5 GeV, which is assumed to be generated after the
gluons in QCD have been integrated out. Its form is motivated by the
sum of one gluon exchange and a linearly rising potential, and it is
assumed to  depend on the variable $t=(p_1-p_3)^2$ only. After
Fierz-transforming the  one-gluon exchange diagram in spin and colour
space and extracting the contribution proportional to the spin, colour,
and flavour structure (\ref{5.3}), this boundary condition reads
\be\label{5.10}
\lambda_\Lambda(p_1,p_2,p_3,p_4)=\frac{2\pi\alpha_s}{t}+
\frac{8\pi\lambda}{t^2}+D(t)\ee Here $\alpha_s$ is the strong
gauge-coupling constant which we take $\alpha_s \sim.3$, for the
string tension we use $\lambda\sim.18\ {\rm GeV}^2$, and $D(t)$ is a
distribution with support at $t=0$ only, which is related to a
constant in the potential \cite{27} and will be given implicitly later.

Solving the evolution equation (\ref{5.9}) numerically - for details
see the next section - we see indeed a pole-like structure as in eq.
(1.5)
 appearing in $\lambda_k$. At the scale $k_\varphi$, where the
collective fields  are introduced, two conditions should be fulfilled:
first, $\lambda_{k_\varphi}$  should approximately factorize in the
form \be\label{5.11}
\lambda_{k_\varphi}(p_1,p_2,p_3,p_4)=g(p_1,p_2)\tilde
G(s)g(p_3,p_4)\ee Second, $\lambda_{k_\varphi}$ and hence
$\Gamma_{4,k_\varphi}$ should not yet  have become extremely large:
$\Gamma_{4}$ appears on the right hand sides of  the evolution
equations for the higher $N$ point functions $\Gamma_N$ with $N > 4$.
If $\Gamma_{4}$ becomes large, these higher $N$ point functions
necessarily become  large as well, and their neglect becomes a very
questionable approximation in this  regime. (Generally, the
introduction of the here neglected suppression factor  $f_{k_\varphi}$
of eq. (4.5) and the use of eq. (4.6) at some scale  $Ak_\varphi >
k_\varphi$ avoid a possible conflict between these two conditions.)
Within the present model we find, however, that both conditions can be
met  simultaneously to a reasonable extend at a scale $k=k_\varphi
\sim .63$ $GeV$  (see sect. (6) and fig. (4)).

As a result of eq. (5.11) $\Gamma_{4,k_\varphi}$ can be written as
\begin{eqnarray}\label{5.12} &&\Gamma_{4,k_\varphi}=\frac{1}{2}\int
\prod_l \left(\frac{d^4p_l}{(2\pi)^4}\right)
(2\pi)^4\delta^4(p_1+p_2-p_3-p_4)\nonumber\\
&&\Bigl\lbrace[\bar\psi^i_a(-p_1)\psi^b_i(p_2)]g(p_1,p_2)\tilde
G(s)g(-p_4,-p_3) [\bar \psi^j_b(p_4)\psi_j^a(-p_3)]\nonumber\\
&&-[\bar\psi^i_a(-p_1)\gamma_5\psi^b_i(p_2)]g(p_1,p_2)\tilde
G(s)g(-p_4,-p_3)[\bar
\psi^j_b(p_4)\gamma_5\psi^a_j(-p_3)]\Bigr\rbrace\nonumber\\
&&+\Gamma_{4,k_\varphi}'.\end{eqnarray} Now we perform the step to
switch to the effective average action for quarks and mesons
$\Gamma_{k_\varphi,0}[\psi,\bar\psi,\sigma]$. Since we only consider
the quark contributions to the evolution of the scalar part of the
effective average action, there is no advantage to introduce a
nonvanishing infrared cutoff $\tilde k$ for the collective fields
$\sigma$ in the present approximation.  We choose composite operators
which read in momentum space \begin{eqnarray}\label{5.13} &&O^{\
b}_a[\psi,\bar\psi;q]=-i\int\frac{d^4p}{(2\pi)^4}g(p,q-p)
\bar\psi^i_a(-p)\psi^b_i(q-p)\nonumber\\
&&O^{(5)b}_a[\psi,\bar\psi;q]=-\int\frac{d^4p}{(2\pi)^4}g(p,q-p)
\bar\psi^i_a(-p) \gamma_5 \psi^b_i(q-p)\end{eqnarray} where the real
function $g(p_1,p_2)$ is normalized \be\label{5.14} g(0,0)=1\ee such
that $\tilde G(s)$ has dimensions $\sim 1/s$. Inserting eq.
(\ref{5.13}) in (4.6),  one obtains \begin{eqnarray}\label{5.15}
&&\Gamma_{k_\varphi,0}[\psi,\bar\psi,\sigma]=\Gamma_{k_\varphi}[\psi,
\bar \psi]\nonumber\\
&&+\frac{1}{2}\int\frac{d^4q}{(2\pi)^4}\Bigl\lbrace O^{\ b}_a(q)\tilde
G(q^2)O_b^{\ a}(-q)+O^ {(5)b}_{\ \ a}(q)\tilde G(q^2)O^{(5)a}_{\ \
b}(-q)\nonumber\\ &&\qquad +\tilde\sigma^{a}_{\ b}(q)\tilde
G^{-1}(q^2)\tilde \sigma_{\ a}^{b}(-q)+\tilde \sigma^{(5)a}_{\quad
b}(q)\tilde G^{-1}(q^2)\tilde\sigma^{(5)a}_{\quad b}(-q)\nonumber\\
&&\qquad-2\tilde\sigma_{\ b}^{a}(q)O_a^{\ b}(-q)
-2\tilde\sigma^{(5)a}_{\quad b}(q) O^{(5)\ b}_{\ a}(-q)
\Bigr\rbrace.\end{eqnarray} It is convenient to define $\sigma$ by
\begin{eqnarray}\label{5.17}
\tilde\sigma&=&\frac{1}{2}(\sigma+\sigma^\dagger), \nonumber\\
\tilde\sigma^{(5)}&=&\frac{i}{2}(\sigma-\sigma^\dagger)\end{eqnarray}
such that the field $\sigma^a_{\ b}$ transforms under chiral flavour
transformations $U_L(N)\times U_R(N)$ as an $(\bar N,N)$ representation
\be\label{5.18} \delta\sigma^a_{\ b}=i(\theta^a_{V\ c}-\theta^a_{A\
c})\sigma^c_{\ b} -i\sigma^a_{\ c}(\theta^{\ c}_{V\ b}+ \theta^{\
c}_{A\ b})\ee After using
$\Gamma_{k_\varphi}[\psi,\bar\psi]=\Gamma_{2,k_\varphi}[\psi,\bar\psi]
+\Gamma_{4,k_\varphi}[\psi,\bar\psi]$ and eq. (\ref{5.12}) for
$\Gamma_{4,k_\varphi}[\psi,\bar\psi]$, the terms quartic in $\psi$ in
$\Gamma_{k_\varphi}[\psi,\bar\psi,\sigma]$ cancel and we are left with
\begin{eqnarray}\label{5.16} &&\Gamma_{k_\varphi}
[\psi,\bar\psi,\sigma]=
\int\frac{d^4q}{(2\pi)^4}\Bigl\lbrace\bar\psi_a^i(q)q\llap/\psi^a_i(q)
+\frac{1}{2}(\sigma^\dagger)^{a}_{\ b}(q)\tilde
G^{-1}(q^2)\sigma^{b}_{\ a}(q) \\
&&+i\int\frac{d^4p}{(2\pi)^4}g(-q,q-p)\bar\psi^i_a(q)
[\frac{1}{2}(1+\gamma_5)\sigma^a_{\ b}(p)
+\frac{1}{2}(1-\gamma_5)(\sigma^\dagger)^a _{\
b}(-p)]\psi^b_i(q-p)\Bigr\rbrace\nonumber\end{eqnarray} We have
neglected here the term $\Gamma_{4k_\varphi}'$ in eq. (\ref{5.12})
which will be dropped in the following.

Except for the momentum dependence of $g$ the effective average action
for composite fields $\Gamma_{k_\varphi}$ describes a standard
fermion-scalar theory with Yukawa coupling but without scalar
self-interactions.

In order to compute $\Gamma_{0,0}[0,\sigma]$, we use the flow equation
for  $\Gamma_{k,0}[0,\sigma]$ in the form of the fermionic version
(\ref{C.5}) of eq.  (2.23) \be\label{5.19}
\partial_t\Gamma_{k,0}[0,\sigma]=-{\rm Tr}\left\lbrace \frac{\partial
R_{kF}}{\partial t}(\Gamma_{k,0}^{(2)}\Bigr\vert_{\psi=0}
+R_{kF})^{-1}\right\rbrace\ee We observe that
$\Gamma^{(2)}_{k,0|\psi=0}$ does not mix the fermionic and  bosonic
parts and we therefore only need to consider the second functional
derivative with respect to the fermions.  Apart from the inverse free
quark propagator the only contribution to  $\Gamma_{k,0}^{\\ \
(2)}|_{\psi=0}$ originates from the Yukawa coupling
$\sim\sigma\bar\psi\psi$, which we assume to be independent of $k$ in
the form specified by (\ref{5.16}). In this approximation  the relevant
part of $\Gamma^{(2)}_{k,0|\psi=0}$ becomes independent of $k$ and eq.
(\ref{5.19}) is easily integrated. The result is  \be\label{5.20}
\Gamma_{k,0}[0,\sigma]=-{\rm
Tr}\ln\left(\frac{\Gamma^{(2)}+R_k}{\Gamma^{(2)}
+R_{k_\varphi}}\right)+\frac{1}{2}\sigma^\dagger\tilde G^{-1}\sigma,\ee
where we implemented the boundary condition from (\ref{5.16})
\be\label{5.21}
\Gamma_{k_\varphi,0}(0,\sigma)=\frac{1}{2}\sigma^\dagger\tilde
G^{-1}\sigma\ee in agreement with eq. (4.6).

The effective potential for the field $\sigma$ or its effective wave
function normalization can now easily be obtained from eq. (\ref{5.20})
for $k=0$:  In the case of the effective potential $\Gamma^{(2)}$ on
the r.h.s. of eq. (\ref{5.20})  has to be evaluated for constant
configurations of the field $\sigma$, and the  momentum dependent
collective field propagator $\tilde G^{-1}(q^2)$ is only  needed at
$q^2 = 0$. In the case of the effective wave function normalization
$Z_{k}$ contributions arise both from an expansion of the  ${\rm
Tr}\ln$ term  up to second order in the external momentum ($O(q^2)$),
and from a corresponding  expansion of $\tilde G^{-1}(q^2)$. Note that
this second contribution is already  present at the scale $k_{\varphi}$
according to eq. (\ref{5.21}); thus an often  used "compositeness
condition" $Z_{k}=0$ \cite{24,25} does not hold at  $k=k_{\varphi}$.

As we will see in the next section, this effective potential exhibits
indeed a  nontrivial minimum for a real diagonal $\sigma$. This
expectation value is invariant under the vectorlike flavour symmetry,
but carries nonvanishing axial charges. Our model therefore describes
dynamical chiral symmetry breaking in this  QCD-motivated fermionic
theory.

\section{Numerical computations and results} \setcounter{equation}{0}

In this section we describe in more detail the procedure outlined in
the last section, and we give an account of the results of the
numerical solution  of the evolution equation. The technique for
integrating numerically the flow equation (\ref{5.9}) follows the one
developed in \cite{7}. The function $\lambda_k$ can only depend on six
independent Lorentz-invariant products of the momenta, which we denote
by \begin{eqnarray}\label{6.1} &&s=(p_2+p_1)^2=(p_3+p_4)^2,\quad
t=(p_1-p_3)^2=(p_2-p_4)^2\nonumber\\ &&v_1=p_1^2,\quad v_2=p^{\
2}_2,\quad v_3=p_3^{\ 2},\quad v_4=p_4^2.\end{eqnarray} Then we switch
from the function $\lambda_k(p_i)$ to its Laplace transform with
respect to the Lorentz five invariants $t, v_1,v_2,v_3,v_4$:
\be\label{6.2}
\lambda_k(p_i)=\int^\infty_0dl_tdl_1dl_2dl_3dl_4C_k(s,l_i)
e^{-l_tt-l_1v_1-l_2v_2-l_3v_3-l_4v_4}\ee After inserting (\ref{6.2})
into (\ref{5.9}) and using  \be\label{6.3}
\frac{e^{-\frac{q^2}{\Lambda^2}}-e^{-\frac{q^2}{k^2}}}{q^2}=\int
^{1/k^2}_{1/\Lambda^2}d\alpha e^{-\alpha q^2}\ee the flow equation for
$C_k(s,l_i)$ can easily be derived, since the $d^4q$ integration in
(\ref{5.9}) becomes Gaussian. The dependence of $C_k(s,l_i)$ on the
variables $l_i$ is then discretized, thus $C_k (s,l_i)$ becomes a
function living on a five-dimensional lattice. The integration  of the
flow equation corresponds to an algorithm for updating the lattice for
each step from $k^2$ to $k^2+\Delta k^2$. The boundary condition for
$C_\Lambda(s,l_i)$, with $\Lambda=1.5$ GeV,  corresponding to
(\ref{5.10}) reads \be\label{6.4} C_\Lambda(s,l_i)=\tilde
C(l_t)\delta(l_1)\delta(l_2)\delta(l_3)\delta(l_4) \ee with
\be\label{6.5} \tilde C(l_t)=8\pi\lambda l_t+2\pi \alpha_s\ee As
discussed in \cite{7}, a vanishing of $\tilde C(l_t)$ for
$l_t>l_{t_{max}}$ corresponds to a vanishing of the potential $V(r)$ in
ordinary space for $r\gg(l_{t_{max}}) ^{1/2}$, and is related to the
distribution $D(t)$ in eq. (\ref{5.10}). With the help of the formulas
in \cite{7} a direct relation between $l_{t_{max}}$ and the depth $d$
of the linear part $\lambda r$ of the potential can be derived:
\be\label{6.6} l_{t_{max}}=\frac{\pi d^2}{4\lambda^2}\ee This finite
value for $l_{t_{max}}$ specifies  implicitly the distribution
 $D(t)$ in (5.10). We will use (6.6) for $l_{t_{max}}$ with $d=1$ GeV.

With these methods and parameters, we compute $C_k(s,l_i)$ on a lattice
with 50 points in the $l_t$ direction, and 10 points in the
$l_1,l_2,l_3$ and $l_4$ directions. For $k=k_\varphi\sim .63$ GeV we
observe that $C_{k_\varphi}(s,l_i)$ approaches a factorized form
\be\label{6.7} C_{k_\varphi}(s,l_i)\simeq\delta(l_t)\tilde
g(l_1,l_2)\tilde G(s)\tilde g(l_3,l_4)\ee and $\tilde G(s)$ becomes
large for $s\to0$.  In fig. 4 we show, as a function of the scale $k,
\tilde G(s=0)$ as a full line. If $C(s,l_i)$ can be written as a
product as on the r.h.s. of  (6.7), a measure  $f$ defined by
$f=[C(s,l_1=1,l_2=1,l_j)\cdot C(s,l_1=0,l_2=0,l_j)] \ /\
[C(s,l_1=1,l_2=0,l_j)\cdot C(s,l_1=0,l_2=1, l_j)]$ should satisfy
$f\simeq 1$. In fig. 4 we also show this measure $f$ as a dotted line.
Thus for $k\to k_\varphi$ $\lambda_k$  indeed develops the pole-like
characteristic for a bound state and we can perform the transition to a
description in terms of a linear $\sigma$-model coupled to quarks.

The next step involves the computation of the effective scalar
potential from (\ref{5.20}). In the present model  $\Gamma^{(2)}$
contains the free fermionic propagator and the Yukawa coupling to
$\sigma^a_{\ b}$ as in eq. (\ref{5.16}): \begin{eqnarray}\label{6.8}
&&(\Gamma^{(2)}(q,q'))^{a}_{\
b}=q\llap/\delta^a_b(2\pi)^4\delta^4(q-q')\\ &&+\frac{i}{2}g(-q,q')
\lbrace\sigma^a_{\ b}(q-q')+(\sigma^\dagger)^a_{\ b}(q'-q)
+(\sigma^a_{\ b}(q-q')-(\sigma^\dagger)^a_{\ b}(q'-q))\gamma_5\rbrace
\nonumber\end{eqnarray} In addition $\Gamma^{(2)}$ is diagonal in
colour space. We will be interested in the effective potential for the
field $\sigma_{\ a}^{b}$. This obtains for a spatially homogeneous
configuration which we will take diagonal and real: \be\label{6.9}
\sigma_{\ a}^{b}(p)=\sigma\delta^b_a(2\pi)^4\delta^4(p)\ee With
(\ref{6.8}) for $\Gamma^{(2)}$ and the form (\ref{5.6}) of $R_{kF}$ the
equation (\ref{5.20}) for the effective potential $V_k(\sigma)$ becomes
\be\label{6.10} V_k(\sigma)=\frac{1}{2}N_f\tilde G^{-1}(0)\sigma^2
-2N_fN_c\int\frac{d^4q}{(2\pi)^4}\ln\frac{P_F^{(\Lambda,k)}
(q)+g^2(-q,q)\sigma^2}
{P_F^{(\Lambda,k_\varphi)}(q)+g^2(-q,q)\sigma^2}\ee with
\be\label{6.11}
P_F^{(\Lambda,k)}(q^2)=\frac{q^2}{(exp({-{q^2}/{\Lambda^2})}
-exp({-{q^2}/{k^2})})^2} \ee

We observe, in particular, the $k$-dependence of the mass term at the
origin; neglecting the $q$-dependence of $g$ we have \be\label{6.12}
\frac{\partial^2 V_k}{\partial\sigma^2}_{|\sigma=0}=N_f\tilde G^{-1}(0)
-\frac{N_fN_c}{8\pi^2}(k^2_\varphi-k^2)\cdot\left
(\frac{4\Lambda^4}{(\Lambda^2+k^2)(\Lambda^2+k^2_\varphi)}-1\right)\ee
which becomes negative for small enough $k$ and  $\tilde G^{-1}(0)$.
This induces spontaneous breaking of the chiral symmetry. For the model
under consideration and the choice of parameters discussed before, the
resulting effective potential exhibits indeed a nontrivial minimum
(independent of $N_f$).  In fig. 5 we show $V_k(\sigma)$ for $k=$ .63,
.27, .18 and 0 GeV.  Numerically the minimum of $V_0(\sigma)$ is found
to be at  \be\label{6.13} \sigma_0=.18\ {\rm GeV}\ee Due to our
convention (\ref{5.14}) for $g(0,0)$, and the $\sigma\bar\psi\psi$
coupling present in $\Gamma[\psi,\bar\psi,\sigma]$, this number can
directly be interpreted as a constituent quark mass.

In order to compute the chiral condensate $<\bar\psi\psi>$ from the
vacuum expectation value $\sigma_0$ we first Fourier-transform the
operator (\ref{5.12}) \begin{eqnarray}\label{6.14} &&O_a^{\
b}(x)=-i\int d^4yd^4z\bar\psi^i_a(x+y)\psi^b_i(x+z)g(y,z)\nonumber\\
&&g(y,z)=\int\frac{d^4q}{(2\pi)^4}\frac{d^4q'}{(2\pi)^4}\exp
i(q_\mu'z^\mu-q_\mu y^\mu)g(-q,q')\end{eqnarray} For constant
$g(-q,q')=1$ the operator $O_a^{\ b}$  reduces to $\bar\psi^i_{\
a}(x)\psi_i ^{\ \beta}(x)$ (up to the phase factor which appears as a
consequence of our Euclidean signature convention). The smoothening due
to the wave function $g(y,z)$ gives a well-defined regularized
meaning\footnote{ An additional smoothening will be related to the
appearance of $f_{k_\varphi}$ in the definition of $O$ (4.4) in a more
accurate treatment.}
 to the chiral condensate $<\bar\psi\psi>$ as the expectation value of
O \be\label{6.15} <\bar\psi\psi>=\frac{1}{N_f}<{\rm Tr}\ O(x)>\ee The
exact relation of our field-theoretical definition of $<\bar\psi\psi>$
to more phenomenological definitions in the context of QCD sum rules
remains to be established. We emphasize that our formalism offers, in
principle, the possibility to provide well-defined field-theoretical
definitions for the concepts underlying QCD sum rules or vacuum
correlators. The relation between the $vev$ of the collective field
$\sigma_a^{\ b}(x)$  and the expectation value of the composite
operator $O_i^{\ j}$ is given by eq. (3.11) \be\label{6.16}
<\sigma_a^{\ b}>=\tilde G(0)<O_a^{\ b}>.\ee Therefore the relation
between $\sigma_0$ and the QCD quark condensate reads \be\label{6.17}
\sigma_0=\tilde G(0)<\bar\psi\psi>.\ee With $\tilde G(0)=33.2$ ${\rm
GeV}^{-2}$ we thus find \be\label{6.18} <\bar\psi\psi>=(175\  {\rm
MeV})^3.\ee It should be clear that the collective fields $\sigma_{\
a}^{b}$ contain the massless pionic degrees of freedom, once they are
expanded around the $vev$ (\ref{6.9}). We can also obtain the pion
decay constant $f_\pi$,  if we expand the effective action
$\Gamma_{0,0}(0,\sigma)$ up to second order in derivatives acting on
$\sigma$. Let us denote the corresponding kinetic terms in
$\Gamma_{k,0}$ by \be\label{6.19}
\frac{1}{2}\int\frac{d^4q}{(2\pi)^4}Z_{k}\sigma^{\dagger a}_{\quad
b}(q)q^2 \sigma_{\ a}^{b}(q);\ee With this normalization we have
\be\label{6.20} f_\pi=\sqrt{2 Z_{k=0}}\cdot \sigma_0\ee As stated below
eq.(\ref{5.21}),  $Z_{0}$ gets contributions from both terms on the
r.h.s. of eq. (\ref{5.20}).  In the model under consideration we
actually find a dominant contribution $.825$ arising from the term
involving $\tilde G^{-1}$; together with a contribution  $.017$ from
the fermionic loop we thus obtain \be\label{6.21} f_\pi=234\ {\rm
MeV}\ee which has to be compared with the experimental value $f_\pi=$
93 MeV. We observe a  substantial discrepancy, but expect that the
dominant contribution arising from the  term involving $\tilde G^{-1}$
is particularly sensitive to the rather crude  approximations employed
in this paper.

\section{Conclusions} \setcounter{equation}{0} We have presented in
this paper a formalism to describe bound states and condensates of
composite operators in terms of an effective average action. The scale
dependence of this effective action is determined by an exact
nonperturbative  evolution equation. The properties of the vacuum and
excitations around it can be infered from the solution of the flow
equation for $k\to 0$. Such a solution allows to extrapolate  from the
short distance physics to the long distance physics.
 We have deviced a detailed  prescription how a change of relevant
variables can be performed in the course of  this evolution.

The practical use of the exact flow equation is closely related to the
existence of a suitable truncation scheme  which permits to solve it
approximately. We have demonstrated the existence of nonperturbative
approximation schemes for a QCD-motivated quark model. Our computations
in sect. 5 and 6 may be viewed as  first steps in a systematic
expansion in 1PI Green functions  with a fixed number of external legs
for the quark fields. Here the truncation neglects many-quark-operators
but retains the full momentum dependence  of the lower 1PI  vertices.
On the other hand the truncation for the meson part of the effective
action may be developed as an expansion in momenta (or appropriate
functions of momenta). This retains the full field dependence on the
zero momentum modes and  can therefore give a detailed picture  of the
meson potential and kinetic term.  Although in the present paper we use
very rough approximations we have demonstrated that our formalism
provides not only a formal tool to describe  phenomena  as complex as
dynamical chiral symmetry breaking, but that suitable truncations of
the $k$-dependent effective action allow  also to compute
phenomenologically interesting numbers.

Here, however, no effort was made to discuss systematically the
dependence of the  results  on various parameters entering the
approximations, like $\Lambda,k_\varphi$ or the string tension
$\lambda$. We defer such an investigation to future work, where less
drastic approximations to the effective action will be made. Such
extensions do  actually not necessarily involve an enormous amount of
additional computational effort. The following parts of such a
programme are certainly feasable and have  partially  already been
performed: In the case of the four point
 function $\Gamma_{4,k}$ more spin, flavour and colour structures can
be taken into account. After the  introduction of the collective fields
the $k$-dependence of the Yukawa-like  couplings  and the quark wave
function normalisation do not have to be neglected.

The meson self-interactions can be included on the r.h.s. of the flow
equation, using the picture with infrared cutoff for both fundamental
and composite fields developed in sect. 4. The transition from the
description in terms of only fundamental  fields to a language with
collective fields  can be smoothened by an appropriate  separation of
the bound-state scale from the scales $k_\varphi, Ak_\varphi$ as
discussed in sect. 5.
 Quark mass terms can be included without any conceptual  difficulties.
We are  also  free to include the $\rho$-mesons as additional
collective fields. Finally, we may replace the ``initial conditions''
for the four quark vertex at the  ultraviolet cutoff scale $\Lambda$ by
a solution of the evolution equation for QCD, i.e. the coupled system
of quarks and gluons. The practical applicability of  the effective
average action for nonperturbative problems in gauge theories has
already been demonstrated \cite{26}.  Although this  program still
needs many steps it seems not impossible to us that it may finally lead
to a reliable computation of chiral  condensates and the meson spectrum
and interactions.

\section*{Appendix A: Green functions for collective fields and
composite operators} \renewcommand{\theequation}{A.\arabic{equation}}
\setcounter{equation}{0} The generating functional
$\Gamma_0[\varphi,\sigma]$ contains all information about Green
functions for  the collective field $\sigma$, and hence as well as for
the composite operator $\tilde O$.  The relation between the two sets
of Green functions is given in this section. The generating functional
for the 1PI Green functions for the collective field obtains from
$\Gamma_0[\varphi,\sigma]$ as \begin{eqnarray}\label{A.1}
\Gamma[\sigma]&=&\Gamma_0[\varphi_0[\sigma],\sigma]\nonumber\\
\frac{\delta\Gamma_0}{\delta\varphi}_{|\varphi_0}&=&0 \end{eqnarray} In
particular, the minimum of $\Gamma_0$ \begin{eqnarray}\label{A.2}
\frac{\delta\Gamma_0}{\delta\sigma}_{|\bar\varphi,\bar\sigma}&=&0
\nonumber\\
\frac{\delta\Gamma_0}{\delta\varphi}_{|\bar\varphi,\bar\sigma}&=&0
\end{eqnarray} determines the expectation values of $\chi$ and $\tilde
O[\chi]$ \begin{eqnarray}\label{A.3}
<\chi^\alpha>&=&\bar\varphi^\alpha\nonumber\\ <\tilde
O[\chi]^i>&=&\bar\sigma^i. \end{eqnarray} The higher correlations for
the composite operator $\tilde O[\chi]$  are not directly given by the
functional derivatives of $\Gamma[\sigma]$ with respect to $\sigma$ but
they are simply related to them. From (\ref{2.1}) it follows
immediately that the  generating functional for the connected Green
functions for $\tilde O$ is \be\label{A.4} \hat
W[J,K]=W_0[J,K]-\frac{1}{2} K^\dagger\tilde G K. \ee Comparing the
Legendre transforms of $\hat W$ and $W_0$ one finds that the
generating functional for the 1PI Green functions for $\tilde O$ is
given by $\hat\Gamma$  \begin{eqnarray}\label{A.5} \hat\Gamma&=&-\hat
W+J^\dagger\varphi+K^\dagger\hat\sigma\nonumber\\
&=&-W_0+\frac{1}{2}K^\dagger\tilde G K+J^\dagger\varphi+K^\dagger
\sigma-K^\dagger\tilde G K\nonumber\\ &=& \Gamma_0-\frac{1}{2}
K^\dagger\tilde G K \end{eqnarray} Here we have used \be\label{A.6}
\hat\sigma^i=\frac{\partial\hat W}{\partial K_i}=\sigma^i- (\tilde G
K)^i \ee The generating functional for $\varphi$ and $\tilde
O[\varphi]$ is therefore obtained from $\Gamma_0$ be straightforward
algebraic manipulations \begin{eqnarray}
\hat\Gamma[\varphi,\hat\sigma]&=&\Gamma_0[\varphi,\sigma]-\frac{1}{2}
\frac{\delta\Gamma_0}{\delta\sigma^i}\tilde G^i_j\frac{\delta
\Gamma_0}{\delta \sigma_j^*}\label{A.7}\\
\hat\sigma^i&=&\sigma^i-\tilde G^i_j\frac{\delta\Gamma_0}{\delta
\sigma_j^*} \label{A.8} \end{eqnarray} As expected, the difference
between $\hat\Gamma$ and $\Gamma_0$ is irrelevant for the expectation
values (\ref{A.3})
 but it accounts for (\ref{2.11}) and generalizations to higher 1PI
vertices. The Green functions for $O$ obtain from the Green functions
for $\tilde O$ by simple rescaling (\ref{2.8a}).

\section*{Appendix B: Exact evolution equation restricted to composite
fields} \renewcommand{\theequation}{B.\arabic{equation}}
\setcounter{equation}{0} If $\varphi_0$ depends on $\sigma$ one has to
correct (\ref{4.3}) according to \be\label{B.1}
\frac{d}{d\sigma}\left(\frac{\partial\Gamma}{\partial\varphi}
[\varphi_0[\sigma],\sigma]\right)=\frac{\partial^2\Gamma}
{\partial\sigma\partial \varphi}[\varphi_0[\sigma],\sigma]
+\frac{\partial^2\Gamma}{\partial\varphi^2}[\varphi_0[\sigma],\sigma]
\frac{d\varphi_0}{d\sigma}=0 \ee and use \be\label{B.2}
\frac{d^2\Gamma[\sigma]}{d\sigma^2}=\frac{\partial^2\Gamma}
{\partial\sigma^2}
[\varphi_0[\sigma],\sigma]+\frac{\partial^2\Gamma}{\partial
\sigma\partial\varphi}
[\varphi_0[\sigma],\sigma]\frac{d\varphi_0}{d\sigma} \ee since
$d/d\sigma=\partial/\partial\sigma+(d\varphi_0/d\sigma)\partial
/\partial\varphi$. The evolution equation (\ref{2.28}) expressing
$\partial\Gamma_ {\tilde k}[\sigma] /\partial\tilde t$ in terms of
$\Gamma_{\tilde k}^{(2)}[\sigma]$ remains nevertheless exact. This can
be seen most easily by using  an implicit functional integral
representation for  $\Gamma_{\tilde k}[\sigma]$
\begin{eqnarray}\label{B.3} &&\exp-\Gamma_{\tilde
k}[\sigma]\nonumber\\ &&=\int{\cal D}\chi{\cal D}\rho\exp-
\left\{S[\chi]+\frac{1}{2}\tilde G^{-1}(\rho-\tilde G O[\chi])^2-
\frac{\delta\Gamma_{\tilde
k}[\sigma]}{\delta\sigma}(\rho-\sigma)+\tilde\Delta_{ \tilde k}
S[\rho-\sigma]\right\}\nonumber\\ &&=\int{\cal D}\rho'\exp-\left\{
S_{eff}[\rho+\rho']-\frac{\delta\Gamma_{\tilde k}
[\sigma]}{\delta\sigma}\rho'+\tilde\Delta_{\tilde k} S[\rho']\right\}.
\end{eqnarray}  We note the complete analogy with the functional
integral representation \cite{16} of the effective average action for
the fundamental fields \be\label{B.4} \exp-\Gamma_k[\varphi]=\int {\cal
D}\chi'\exp-\left\{ S[\varphi+\chi']
-\frac{\delta\Gamma_k[\varphi]}{\delta\varphi}\chi'+\Delta_k
S[\chi']\right\}. \ee The exact evolution equation can be infered  from
this analogy by observing that the precise  functional form of
$S_{eff}$ needs not to be known.  Alternatively, one can derive the
evolution equation starting directly from (\ref{B.4}) and following the
procedure of ref. \cite{11}. One arrives at \be\label{B.5}
\frac{\partial}{\partial\tilde t}\Gamma_{\tilde k}[\sigma]=\frac{1}{2}
\Tr\left\{(\Gamma_{\tilde k}^{(2)}[\sigma]+\tilde R_{\tilde k})^{-1}
\frac{\partial}{\partial\tilde t}\tilde R_{\tilde k}\right\} \ee with
$\Gamma^{(2)}_{\tilde k}$ now denoting second functional derivatives
with respect  to $\sigma$.

\section*{Appendix C: Effective average action with fermions}
\renewcommand{\theequation}{C.\arabic{equation}}
\setcounter{equation}{0} In this short section we present the
generalization of our formalism for fermions. We write the fermionic
infrared cutoff term as \be\label{C.1} \Delta_k
S_F=\left(\frac{1}{2}\right)\bar\eta_\alpha(R_{kF})^\alpha_{\
\beta}\eta^\beta. \ee Here $\eta$ are Grassmann variables which are
either Dirac spinors or Weyl spinors in even dimensions, or Majorana or
Majorana-Weyl spinors in those dimensions where this is possible {}.
(We use the conventions for Euclidean spinors of ref. \cite{22}.) The
factor $\frac{1}{2}$ appears only if $\eta$ and $\bar\eta$ are related,
as for the case of Majorana and Majorana-Weyl spinors or a
representation of Weyl spinors in the form of $2^{\frac{d}{2}}$
component spinors with identified components. For Dirac spinors or the
standard  $2^{\frac{d}{2}-1}$ component Weyl spinors this factor is
absent. The formalism of ref. \cite{4}  applies identically for
fermions except for one important minus  sign. First we observe that
the infrared cutoff does not mix bosons and  fermions. The evolution of
the effective average action $\partial\Gamma_k/\partial t$ obtains
therefore separate contributions from bosons and fermions. Both for the
bosonic and  fermionic contribution one arrives at the identity
(\ref{2.18},\ref{2.19}) \begin{eqnarray}\label{C.2}
\frac{\partial}{\partial
t}\Gamma_k&=&\frac{1}{2}\frac{\partial}{\partial t} (R_{kB})^\alpha_{\
\beta}(<\chi_\alpha^*\chi^\beta>-<\chi_\alpha^*><\chi^\beta>)
\nonumber\\ &&+\left(\frac{1}{2}\right)\frac{\partial}{\partial
t}(R_{kF})^\alpha_{\ \beta}
(<\bar\eta_\alpha\eta^\beta>-<\bar\eta_\alpha><\eta^\beta>).
\end{eqnarray} The minus sign for the fermionic contribution appears
due to the anticommutation  properties of the Grassmann variables
\be\label{C.3} <\bar\eta_\alpha\eta^\beta>-<\bar\eta_\alpha>
<\eta^\beta>=-(<\eta^\beta\bar\eta_\alpha>-<\eta^\beta>
<\bar\eta_\alpha>)= -((\Gamma^{(2)}_k+R_k)^{-1}_{FF})^\beta_{\
\alpha}\ee Here $\Gamma_k^{(2)}$ and $R_k$ are considered as enlarged
matrices acting  both on bosonic and fermionic degrees of freedom, i.e.
\begin{eqnarray}\label{C.4}
&&(\Gamma_k^{(2)})^\beta_{FF\alpha}=-\frac{\delta^2\Gamma_k}
{\delta\bar \eta_\beta\delta\eta^\alpha}=\frac{\delta^2\Gamma_k}
{\delta\eta^\alpha\delta\bar\eta_\beta}\nonumber\\
&&(\Gamma_k^{(2)})^\beta_{BF\alpha}=\frac{\delta^2\Gamma_k}{\delta
\chi^*_\beta\delta\eta^\alpha}\nonumber\\
&&(\Gamma_k^{(2)})^\beta_{FB\alpha}=\frac{\delta^2\Gamma_k}{\delta
\bar\eta_\beta\delta\chi^\alpha}\nonumber\\
&&(\Gamma_k^{(2)})^\beta_{BB\alpha}=\frac{\delta^2\Gamma_k}{\delta
\chi^*_\beta\delta\chi^\alpha}\end{eqnarray} (We observe that the
second derivatives (\ref{C.4}) are sufficient to describe
$\Gamma_k^{(2)}$ only if $\chi,\chi^*$ and $\eta,\bar\eta$ are not
independent, e.g. for real scalar fields and Majorana spinors. For
Dirac spinors $\eta$ and $\bar\eta$ lead to independent entries in the
matrix of second functional derivatives and one needs, for example,
$\delta^2\Gamma_k/\delta \chi^*_\beta\delta\bar\eta_\alpha$.) The
inclusion of possible fermionic composite operators is straightforward
and our final evolution equation is \be\label{C.5}
\frac{\partial}{\partial t}\Gamma_{k,\tilde k} =\frac{1}{2}{\rm
Tr}\left\lbrace\frac{\partial R_{kB}}{\partial t}
\left(\Gamma_{k,\tilde k}^{(2)}+R_k\right)^{-1}\right\rbrace
-\frac{1}{2}{\rm Tr}\left\lbrace\frac{\partial R_{kF}}{\partial
t}\left(\Gamma_{k,\tilde k}^{(2)}+R_k\right)^{-1}\right\rbrace\ee The
first trace effectively runs only over bosonic degrees of freedom and
the second over fermionic ones. The second functional derivative
$\Gamma_k^{(2)}$ involves both fundamental and composite fields. The
evolution with respect to $\tilde k$ is similar (cf. (\ref{2.28})).

\bigskip \section*{Figure Captions} \begin{description} \item{Fig. 1:}
Vertex joining the collective field $\rho$ to two fundamental fields
$\chi$ as present in the action $S_k[\chi,\rho]$ of eq. (3.3).
\bigskip \item{Fig. 2:}  Form of the term $\sim
O^{\dagger}[\chi]\tilde G O[\chi]$ in the action  $S_k[\chi,\rho]$ of
eq. (3.3).  \bigskip \item{Fig. 3:} Diagrammatic representation of the
evolution equation (5.9) for the four point  function $\lambda_k$.
\bigskip \item{Fig. 4:}  Full line: The collective field propagator
$\tilde G(s)$, at $s=0$, as a  function of the scale $k$, in units of
$100$ $GeV^{-2}$.  This result has been obtained from a numerical
computation of $C_{k}$ as described in sect.6, and using the relation
(6.7) between $C_{k}$ and $\tilde G$.\\ Dotted line: The measure $f$,
as defined below eq. (6.7), as a function of the  scale $k$. The
approach of $f$ towards $1$ indicates that the momentum dependence  of
the four point function $\lambda_k$ or its Laplace transform $C_{k}$
indeed  factorizes as used in eq. (6.7).\\ The scale $k_{\varphi}$,
where the collective field is introduced, is chosen at  $.63$ GeV.
\bigskip \item{Fig. 5:} Effective potential for the collective field
$\sigma$, as given by eq.(6.10),  for different values of the infrared
cutoff $k$. The dotted lines correspond to,  from inside to outside,
$k=.63$ $GeV$, $k=.27$ $GeV$ and $k=.18$ $GeV$.  The full line is the
final result for $k=0$. \end{description}

\end{document}